\documentclass[useAMS,usenatbib]{mn2e} 
\newcommand{\FIG}[0]{}

\usepackage{txfonts}
\usepackage{graphicx}
\usepackage{natbib}

\def\lapp{\mathbin{\raise2pt \hbox{$<$} \hskip-9pt \lower4pt \hbox{$\sim$}}}
\def\gapp{\mathbin{\raise2pt \hbox{$>$} \hskip-9pt \lower4pt \hbox{$\sim$}}}
\usepackage{txfonts}

\title[AMRVAC and Relativistic Hydrodynamic simulations for GRB afterglow
  phases]{AMRVAC and Relativistic Hydrodynamic simulations for GRB
  afterglow phases}

\author[Z. Meliani, R. Keppens, F. Casse and
    D. Giannios]{Zakaria Meliani$^{1,2}$\thanks{E-mail: meliani@rijnh.nl}, Rony Keppens$^{1,3,4}$, Fabien Casse$^{5}$
    and Dimitrios Giannios$^{2}$\\
$^1$ FOM-Institute for Plasma Physics Rijnhuizen P.O. Box 1207 3430 BE Nieuwegein, Netherlands\\
$^2$ Max Planck Institute for Astrophysics, Box 1317, D-85741 Garching
  Germany\\
$^3$ Centre for Plasma Astrophysics, K.U.Leuven, Belgium\\
$^4$ Sterrenkundig Instituut, Utrecht, Netherlands\\
$^5$ AstroParticule \& Cosmologie (APC)\\
  UMR 7164 CNRS - Universit\'e Paris 7,
    10 rue Alice Domon et L\'eonie Duquet 75025 Paris Cedex 13, France\\}

\begin{document}

\date{Accepted;  Received }

\pagerange{\pageref{firstpage}--\pageref{lastpage}} \pubyear{2007}
\maketitle

\begin{abstract}
We apply a novel adaptive mesh refinement code, AMRVAC, to numerically investigate the various
evolutionary phases in the interaction of a relativistic shell with its
surrounding cold Interstellar Medium (ISM). We do this for both 1D isotropic as well as full 2D
jetlike fireball models.
This is relevant for Gamma Ray Bursts, and we demonstrate that, thanks to
the AMR strategy, we resolve the internal structure
of the shocked shell-ISM matter, which will leave its imprint on the GRB afterglow.
We determine the deceleration from an initial
Lorentz factor $\gamma=100$  up to the  almost Newtonian
$\gamma\sim{\cal O}(2)$ phase of the flow.
We present axisymmetric 2D shell evolutions, with the 2D extent characterized by their initial opening angle.
In such jetlike GRB models, we discuss the differences with the 1D isotropic GRB equivalents. These are
mainly due to thermally induced sideways expansions of both the
shocked shell and shocked ISM regions. We found that the propagating 2D ultrarelativistic
shell does not accrete all
the surrounding medium located within its initial opening angle. Part of this ISM matter gets pushed away
laterally and forms a wide bow-shock configuration with swirling flow patterns trailing the thin shell.
The resulting shell deceleration is quite different from that found in isotropic GRB
models. As long as the lateral shell expansion is merely due to ballistic spreading of
the shell, isotropic and 2D models agree perfectly. As thermally induced expansions
eventually lead to significantly higher lateral speeds, the 2D shell interacts with comparably more
ISM matter and decelerates earlier than its isotropic counterpart.
\end{abstract}

\section{Introduction}

Many high energy astrophysical phenomena involve relativistic flows and shocks.
For example, relativistic flows are invoked
to explain the observed properties of various compact astrophysical objects
\citep{Arons04, Ferrari98, Corbel04}.
Astrophysical relativistic flows can reach a Lorentz
factor of $2-10$ in association with jets from Seyfert and radio loud galaxies
\citep{Pineretal03}, or even go up to Lorentz factors
$10^2-10^3$ for Gamma Ray Burst (GRB) scenarios
\citep{Sari&Piran99, Soderberg&Ramirez-Ruiz01, Meszaros06}.
In the last decade, continued development of numerical algorithms and the
increase in computer power have allowed to significantly progress in
high-resolution, hydrodynamic
numerical simulations in both special and general relativity
(see Marti \& M\~{u}ller 2003).
The enormous time and length scale ranges associated with violent astrophysical
phenomena in relativistic hydrodynamics (RHD), make  Adaptive Mesh Refinement
(AMR) an important algorithmic ingredient for computationally affordable
simulations. RHD numerical simulations, particularly when combined with AMR
capabilities, can investigate many details of
relativistic flow regimes relevant for astrophysics.

In this paper, we concentrate on relativistic dynamics in the fireball model for
the afterglow phases of GRBs, in one and two dimensional simulations.
Since the follow-up detection of GRBs in X-ray \citep{Costa97} and their afterglows at longer wavelengths \citep{Sahu97,
vanParadijsetal97, Galama97, Frail97, Piroetal98}, the cosmological
origin of GRBs has been established \citep{Metzger97, Wijers97}.
These detections confirmed the predictions from the fireball theoretical
model~\citep{Rhoads93, Katz94, Meszaros&Rees97, Vietri97}.
 In this model, a compact source releases a large amount of
energy in a very short timescale, producing a fireball expanding with
relativistic velocity. Its  internal energy gets fully converted to kinetic
energy, leading to a shell expanding with very high Lorentz factor. This cold
shell continues to expand and interact with the circumburst medium, producing a
relativistic shock-dominated evolution. As the shell sweeps up the matter, it
begins to decelerate. Here, we investigate the details of such propagating
relativistic shells with the relativistic hydrodynamics code AMRVAC
\citep{Bergmansetal04}. The AMRVAC code \citep{Keppensetal03} is here for
the first time applied to the numerically challenging regime of high Lorentz
factor, and we therefore include a variety of test problems, demonstrating the
robustness as well as the limitations of our computational strategy.

Up till recently, analysis of GRB flows have largely been done
analytically \citep{Shemi&Piran90, Sari&Piran95, Meszaros&Rees97,
Chiang&Dermer99}, combined with numerical approaches usually
employing a Lagrangian code. These latter works mainly investigate
spherically symmetric GRB scenarios for obvious computational
convenience \citep{Panaitescuetal97, Kobayashietal99,
Kobayashi&Sari00}. Recently, some analytical works started to
investigate the multidimensional jet structure in GRBs
\citep{Rhoads97, Rhoads99, Panaitescu&Meszaros99, Sarietal99,
Kumar&Panaitescu00, Panaitescu&Kumar03, Chengetal01, Orenetal04,
Kumar&Granot03}, and some numerical simulations emerged as well, but
restricted to relatively low (order 25) Lorentz
factor~\citep{Granot01, Cannizzoetal04}.
{Higher speeds were obtained in the numerical simulation of the
propagation of an axisymmetric jet through a collapsing rotating massive star,
as investigated by \citet{Aloy00} to analyse the first phase of GRBs.
In these simulations, the jet is further followed after breakout to a maximum Lorentz
factor of $\gamma_{\rm max}\sim 44$, which is still relatively small to the
values required for GRBs by the fireball model.}
  Therefore, an important area of current
investigations in GRB context is to model the dynamics of narrow
jets of ultra-relativistically flowing ejecta. This is motivated by the
need to reduce the total amount of energy released in GRBs, by
assuming these jets to point towards the observer, as compared to
fully isotropic equivalents. This need is particularly clear for the
exemplary cases of GRB 990123 \citep{Kulkarni99}, GRB 050820A
\citep{Cenkoelal06}, and GRB 050904 \citep{Frailetal06}.

 The detection of polarization
\citep{Covinoetal99, Wijersetal99, Greineretal03, Lazzatietal04} gave
further support to the jetlike model. Evidence for narrow
collimated outflows in GRBs is sustained also by the achromatic
breaks in the afterglow light curves which was predicted analytically (Rhoads (1997,
1999); Sari et al. (1999)) and then observed in a large number of GRBs
\citep{Staneketal99, Sarietal99, Berger00, Panaitescu05,
Gorosabeletal06, Barthelmy05}. In \citet{Bloometal03}, various GRBs were analysed and
in 16 of them, the combination of these breaks in
the spectrum and the jet-like model was used to deduce their effective
energy, which was about $E\sim 10^{51} {\rm ergs}$.
The half  opening angle of such jets in GRBs is inferred to be of order
few degrees. As a result, the afterglow
producing shocked region is collimated too, with a  similar initial opening angle
\citep{Frailetal01, Bergeretal03, Cenkoelal06,Panaitescu05,
Dar&Rujula04}.  In our 2D
simulations, we will concentrate on the afterglow phases in the GRB
evolution starting from  collimated ejecta, and discuss
those dynamical effects causing opening angle changes in
detail. Direct comparison with the evolution of an equivalent 1D spherical shell
is enlightening in this respect.

This paper is organised as follows. We start by reviewing the relativistic
hydrodynamic equations. In Section~\ref{test}, we include several tests to demonstrate the AMRVAC code potential for
realistic RHD computations. In Section~\ref{after}, we present our main
astrophysical application to GRB flows in 1D and 2D models.
\section{Relativistic hydrodynamic equations}\label{rhd}
The special relativistic hydrodynamic evolution of a perfect fluid is governed
by the conservation of the number of particles, and energy-momentum conservation. These two conservation laws can be written as
\begin{eqnarray}
\left(\rho\,u^{\mu}\right)_{\mu}=0\,,\;\;\;
\left(T^{\mu\nu}\right)_{\mu}=0\,.
\end{eqnarray}
where $\rho$, $\vec{u}=\left(\gamma,\gamma\,\vec{v}\right)$, and
$T^{\mu\nu}=\rho\,h\,u^{\mu}\,u^{\nu}+\,p\,g^{\mu\nu}$ define, respectively, the proper
density, the four-velocity and the stress-energy tensor of the perfect fluid. Their definition involves
the Lorentz factor $\gamma$, the fluid pressure $p$, and the relativistic specific enthalpy
$h=1+e+p/\rho$ where
$e$ is the specific internal energy. For the (inverse) metric
$g^{\mu\nu}$, we take the Minkowski metric. Units are taken where the light speed equals unity.

These equations can be written in conservative form involving the Cartesian coordinate axes and the time axis of a fixed
`lab' Lorentzian reference frame as
\begin{eqnarray}
\frac{\partial U}{\partial t} +
\sum^{3}_{j=1}\frac{\partial F^{j}}{\partial x^{j}}=0\,.
\end{eqnarray}
The conserved variables can be taken as
\begin{equation}
U=\left[D=\gamma\,\rho,\vec{S}=\gamma^2\rho\,h\,\vec{v},
\tau=\gamma^2\rho\,h-p-\gamma\rho\right]^{T}\,,
\end{equation}
and the fluxes are then given by
\begin{equation}
F=\left[\rho\gamma\,\vec{v},\gamma^2\rho\,h\,\vec{v} \vec{v}+p\,{\bf I},
    \gamma^2\rho\,h\vec{v}- \gamma \rho\, \vec{v}\right]^{T}\,,
\end{equation}
where ${\bf I}$ is the $3\times 3$ identity matrix.
To close this system of equations, we use the equation of state (EOS) for an ideal
gas, which is the polytropic equation with the polytropic index $\Gamma$,
\begin{equation}
p\,=\,\left(\Gamma - 1\right)\,\rho \,e\,.
\end{equation}

At each time step in the numerical integration, the primitive variables $(\rho,\vec{v},p)$ involved in flux expressions
should
be derived from the conservative variables $U$ resulting in a system of
nonlinear equations.  One can bring this system into a single equation for the pressure
$p$,
\begin{equation}\label{NR}
\tau+D-\gamma(p)\,D-\frac{p+\Gamma p (\gamma(p)^2-1)}{\Gamma -1}=0\,,
\end{equation}
which, once solved for $p$ yields $\vec{v}=\frac{\vec{S}}{\tau+p+D}$.
This nonlinear equation~(\ref{NR}) is solved using a
Newton-Raphson algorithm.

\begin{figure}
\begin{center}
\FIG{
{\rotatebox{0}{\resizebox{7cm}{5cm}{\includegraphics{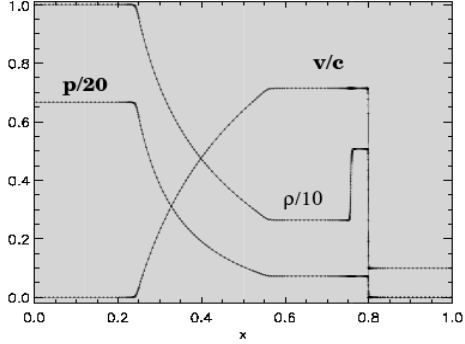}}}}
} \caption{One-dimensional relativistic shock problem in planar
geometry at $t = 0.36$. The solid lines are the analytical solution.}
\label{LabelFig1}
\end{center}
\end{figure}
\section{Testing AMRVAC}\label{test}
In view of the challenges in the numerical investigation of
relativistic fluids, we include here several substantial test
results for code validation. We performed a large series of tests,
some of them shown in this section. An important subclass of test
cases is formed by Riemann problems, whose numerical solution can be
compared to analytical solutions. Other, 2D tests shown here have no
known analytical solution. Therefore, we compare the results of our
simulations with similar results previously obtained by other codes
as documented in the astrophysical literature.

{ The Adaptive Mesh Refinement version of the Versatile Advection Code
(AMRVAC) is specifically designed for
simulating dynamics governed by a system of (near-)conservation laws. Available equations are the Euler and magnetohydrodynamic systems, in both classical and special relativistic versions. The discretization is finite volume based, and various shock-capturing algorithms can be used. The automated AMR
strategies implemented vary from the original patch-based to a novel hybrid block-based approach \citep{holstkep06}. These procedures generate or
destruct hierarchically nested grids with subsequently finer mesh
spacing. The refinement criterion used in AMRVAC is based on a Richardson type
error estimator \citep{Keppensetal03}. In all following
tests, we use a refinement ratio of $2$ between consecutive levels, unless stated otherwise.}

\subsection{One-dimensional test problems}
\subsubsection{Riemann problems}
In 1D Riemann problems, we follow the evolution of an initial
discontinuity between two constant thermodynamical states. In 1D RHD, we then typically find the appearance of up to
three nonlinear waves. Generally, one finds a shock wave
propagating into the lower density/pressure medium, a rarefaction wave propagating at
sound speed into the denser medium, and between these two states, there can be a contact discontinuity.
In the tests that follow, calculations are done in Cartesian geometry on a spatial domain $0\le x\le 1$.
The exact solutions for Riemann problems in relativistic hydrodynamics are discussed for vanishing tangential speed (i.e. $y$ or/and $z$
components for velocities) in \citet{Marti&Muller1994} and
for arbitrary  tangential flow velocity in \citet{Ponsetal00}.

In a first, mild test, we assume an ideal gas with
polytropic index $5/3$ and initial constant
states characterized by $p_{\rm L} =13.3$, $\rho_{\rm L} = 10.0$ (left) and
$p_{\rm R} = 0.66 \times 10^{-6}$, $\rho_{\rm R} = 1.0$ (right),  separated at
the location $x=0.5$. The results at
$t=0.36$ are shown in Fig.~\ref{LabelFig1} with a resolution of  $100$
cells on the base level and $4$ levels, where we also overplot the exact solution.
In the table 1, we present the $L_1=\Sigma \left(\Delta
x_j\right)|\rho_j-\rho(x_j)|$ norm errors of the density $\rho$, where
$\rho(x_j)$ is the exact solution. The accuracy of our result is
comparable to that of \citet{Lucas-Serranoetal04, Zhang&MacFadyen06}.
\begin{table}
\caption{$L_{1}$ errors of the density for the 1D Riemann problem 1 with uniform grid shown at
$t=0.36$}              
\label{table:1}      
\centering                                      
\begin{tabular}{c c}          
\hline\hline                        
Number of grid points & $L_{1}$  \\    
\hline                                   
    200 & $1.15\times 10^{-1}$ \\      
    400 &  $6.4\times 10^{-2}$\\
    800 & $3.2\times 10^{-2}$ \\
    1600 & $1.9\times 10^{-2}$ \\
    3200 & $ 1.06\times10^{-2}$ \\
\hline                                             
\end{tabular}
\end{table}

\begin{figure*}
\begin{center}
\FIG{
{\rotatebox{0}{\resizebox{5cm}{3cm}{\includegraphics{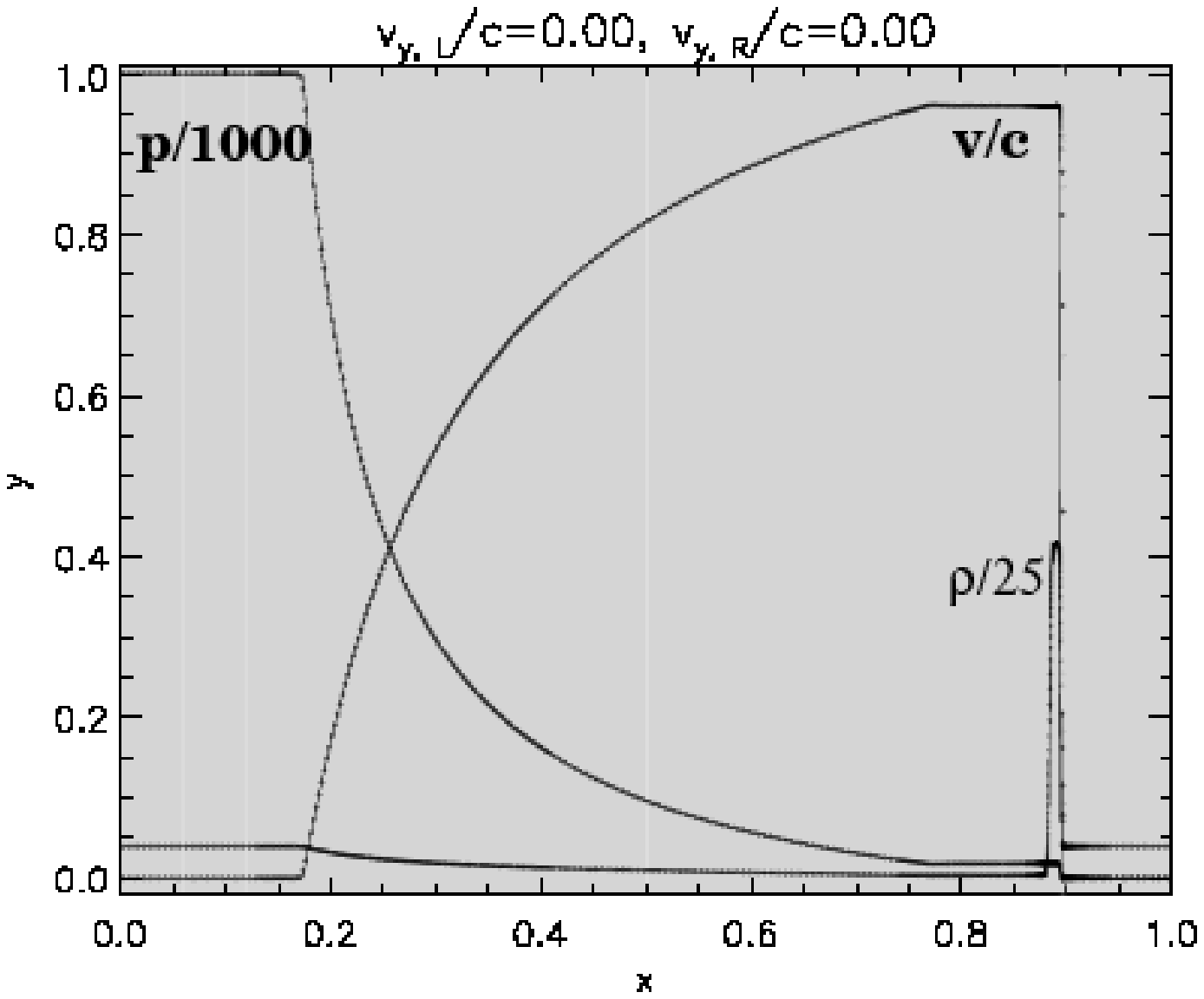}}}}
{\rotatebox{0}{\resizebox{5cm}{3cm}{\includegraphics{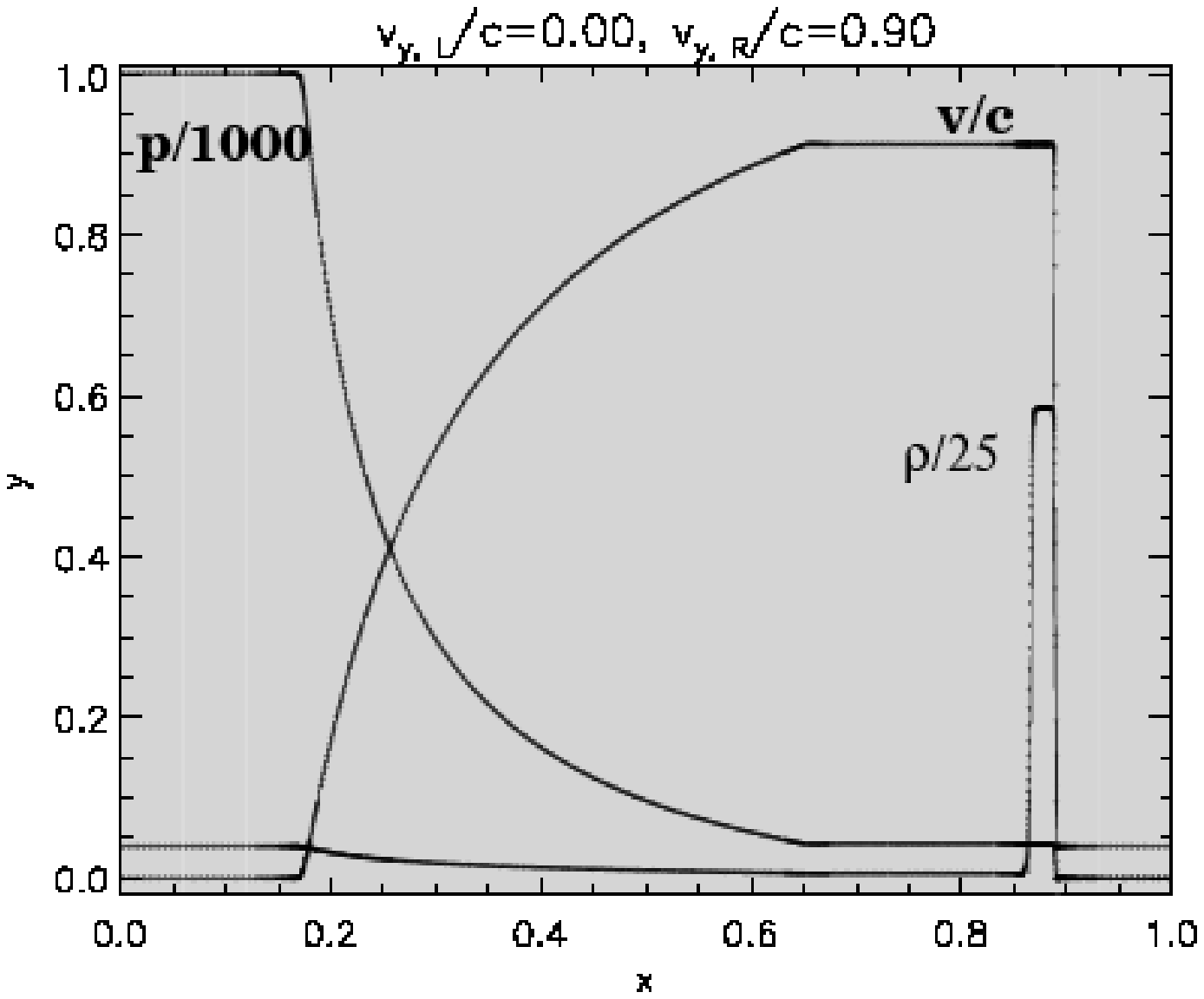}}}}
{\rotatebox{0}{\resizebox{5cm}{3cm}{\includegraphics{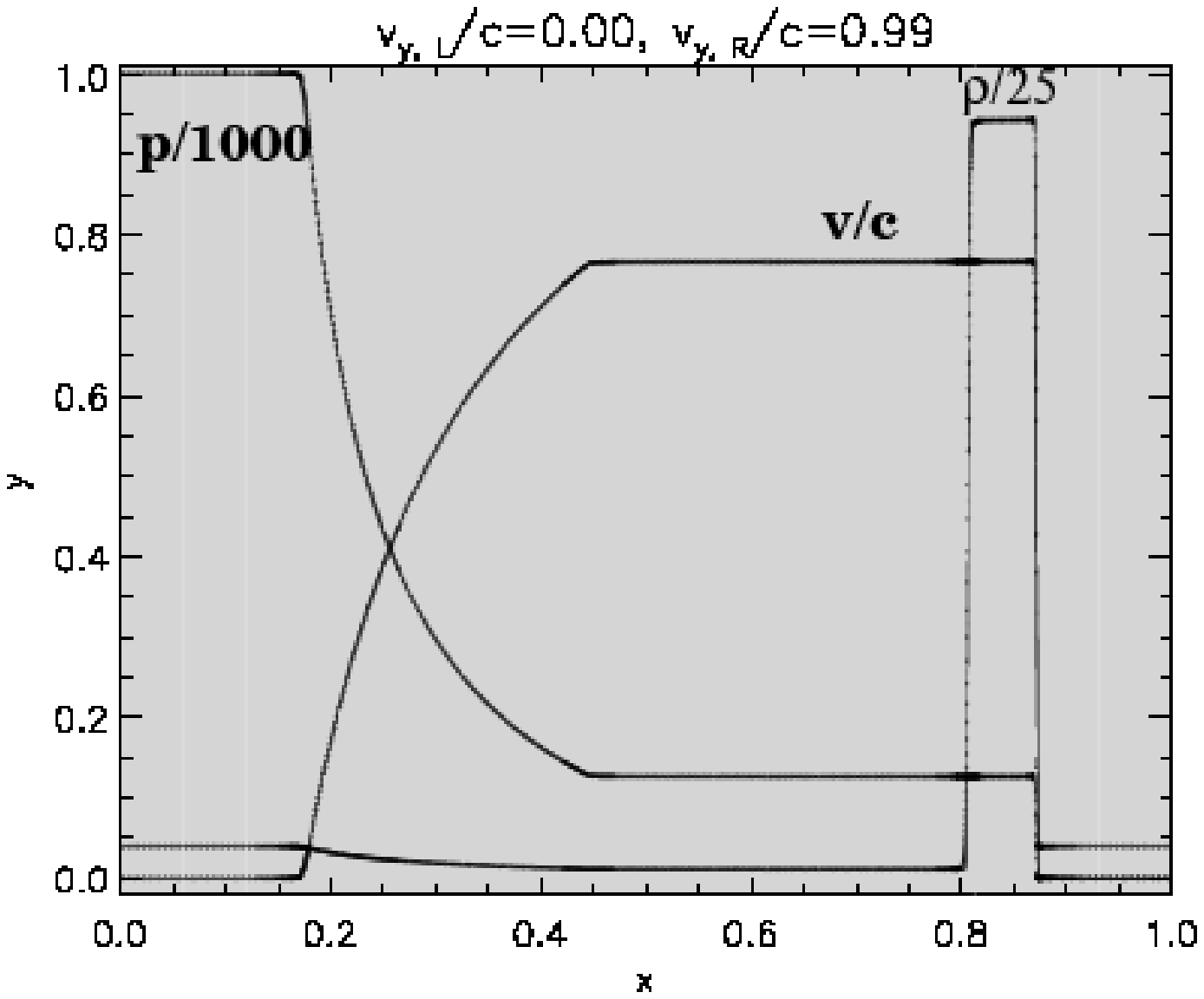}}}}
{\rotatebox{0}{\resizebox{5cm}{3cm}{\includegraphics{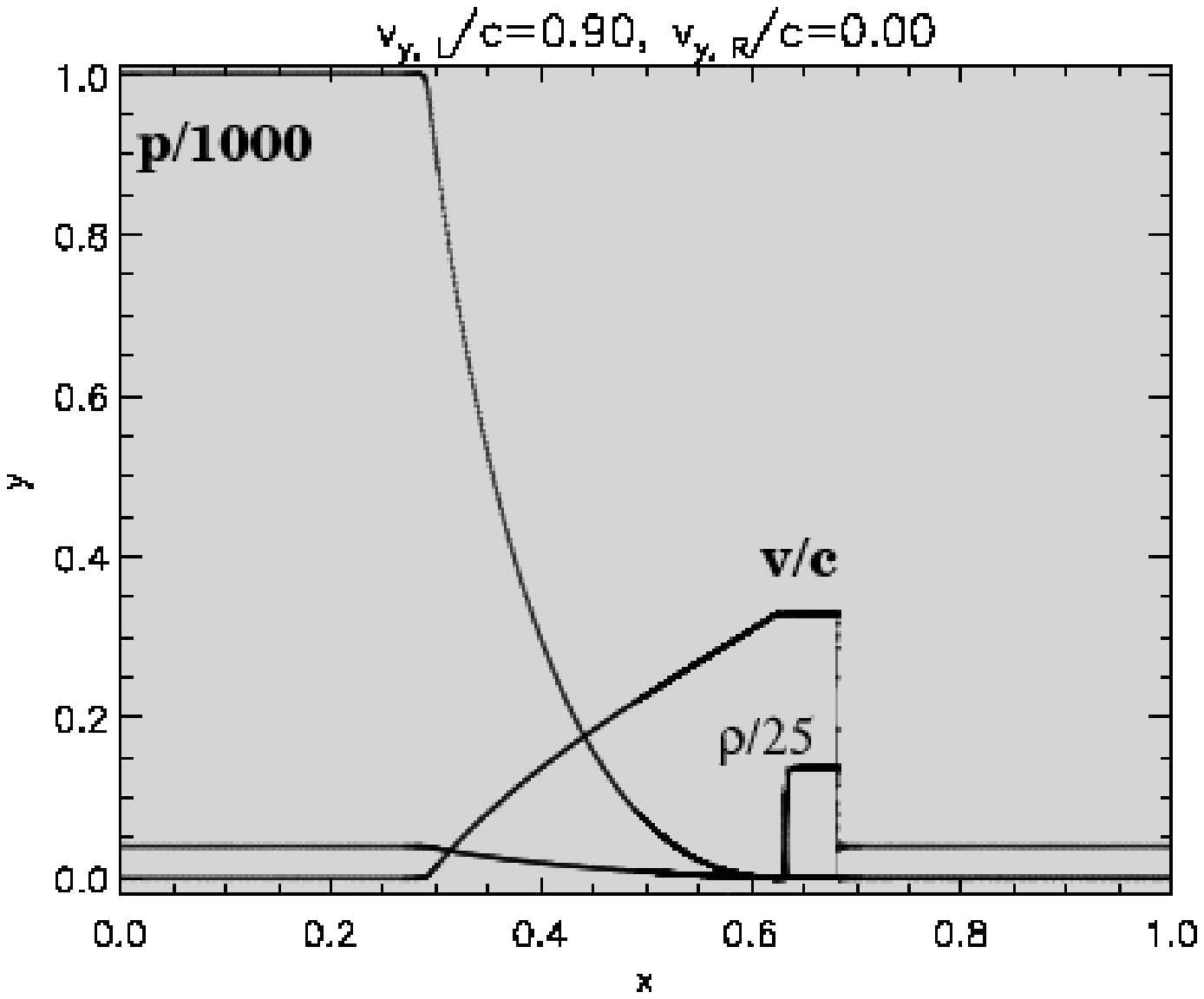}}}}
{\rotatebox{0}{\resizebox{5cm}{3cm}{\includegraphics{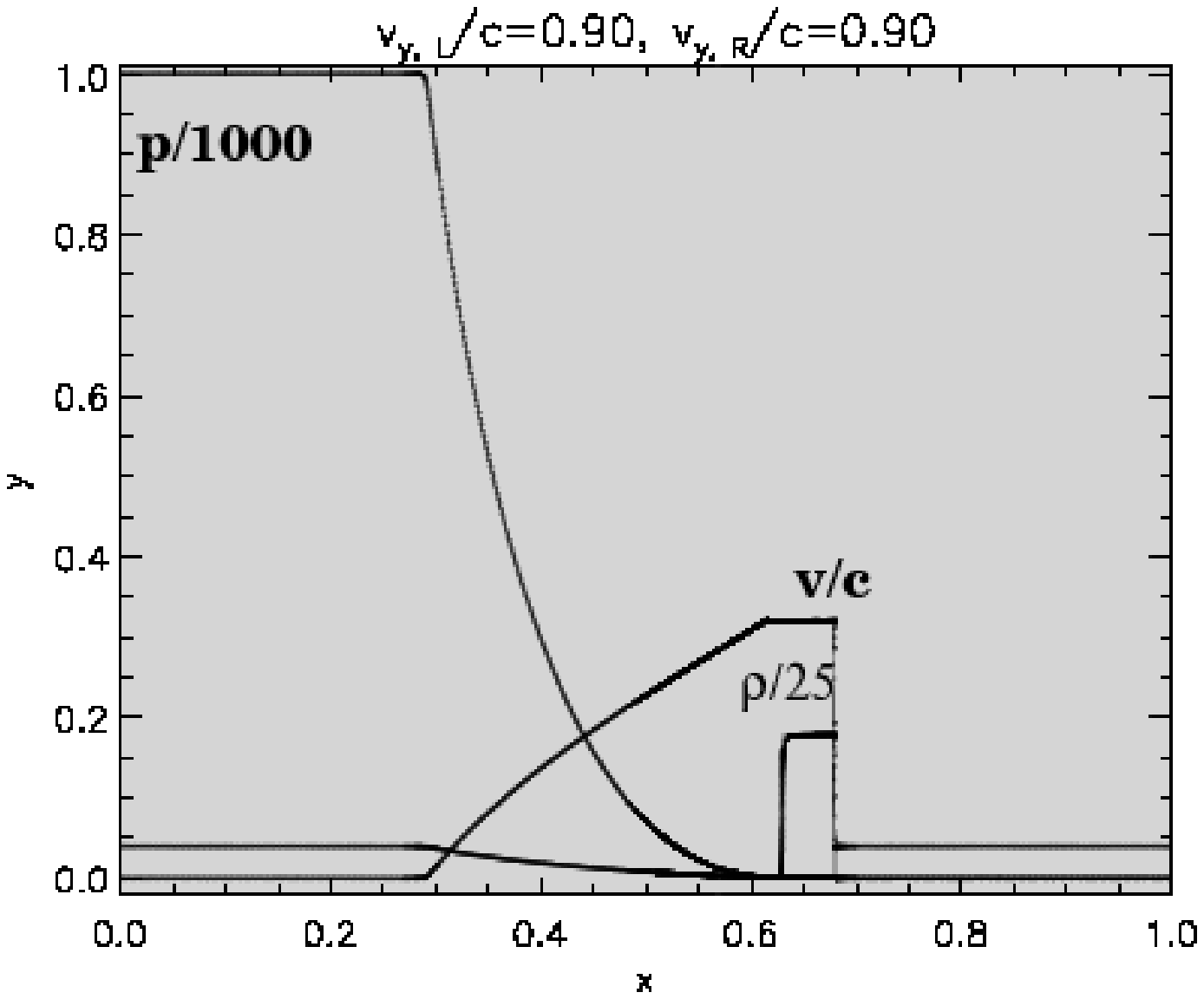}}}}
{\rotatebox{0}{\resizebox{5cm}{3cm}{\includegraphics{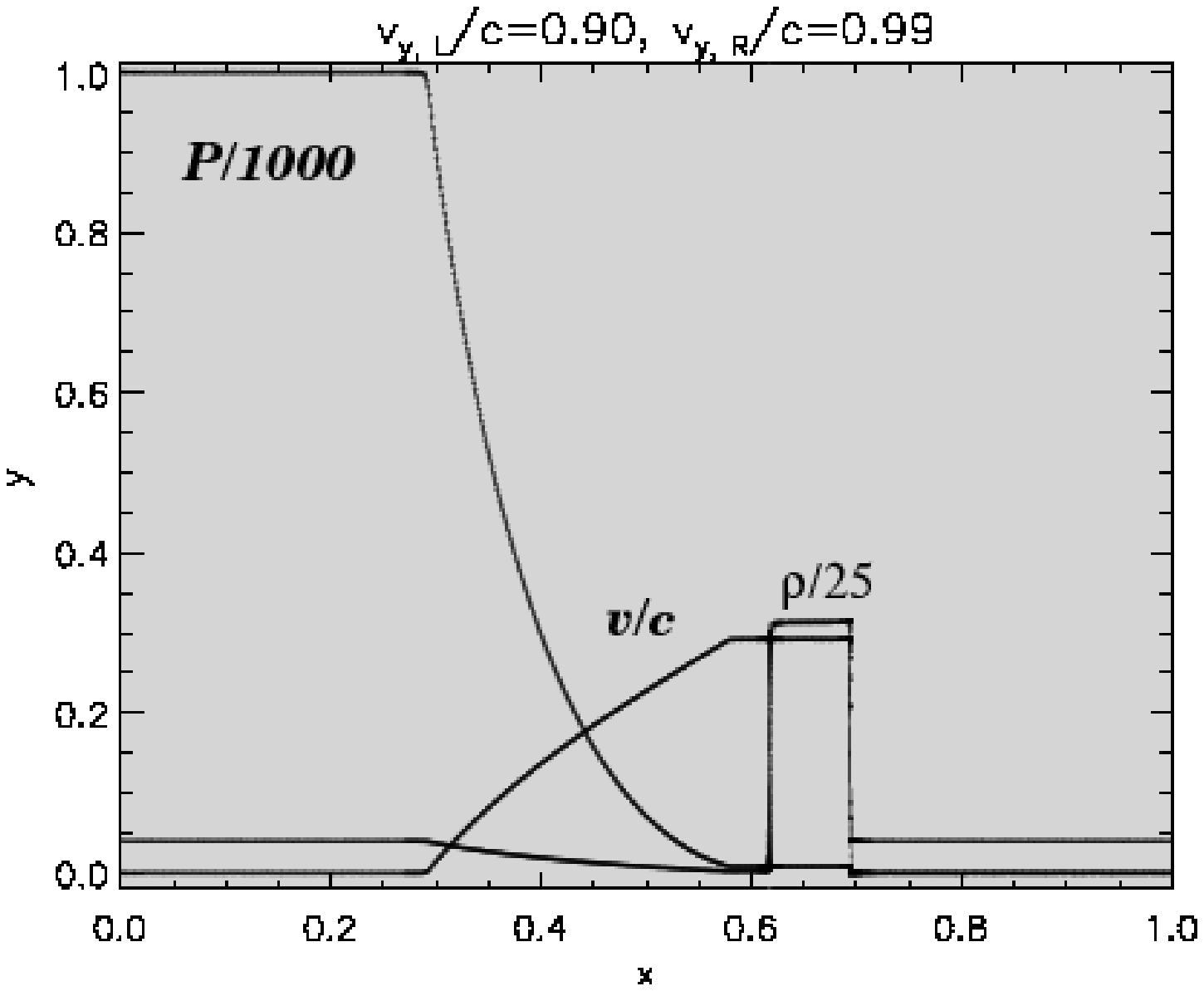}}}}
{\rotatebox{0}{\resizebox{5cm}{3cm}{\includegraphics{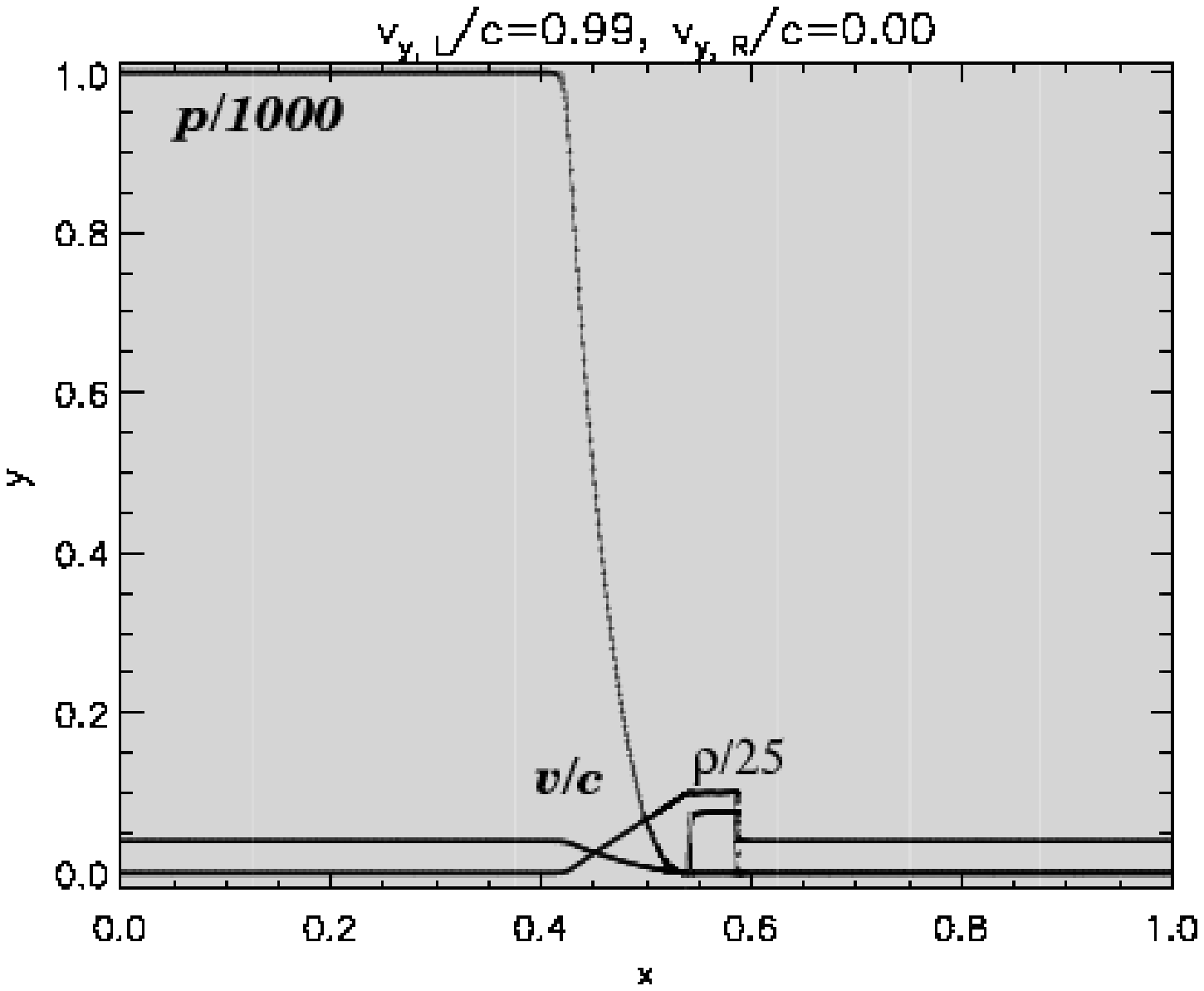}}}}
{\rotatebox{0}{\resizebox{5cm}{3cm}{\includegraphics{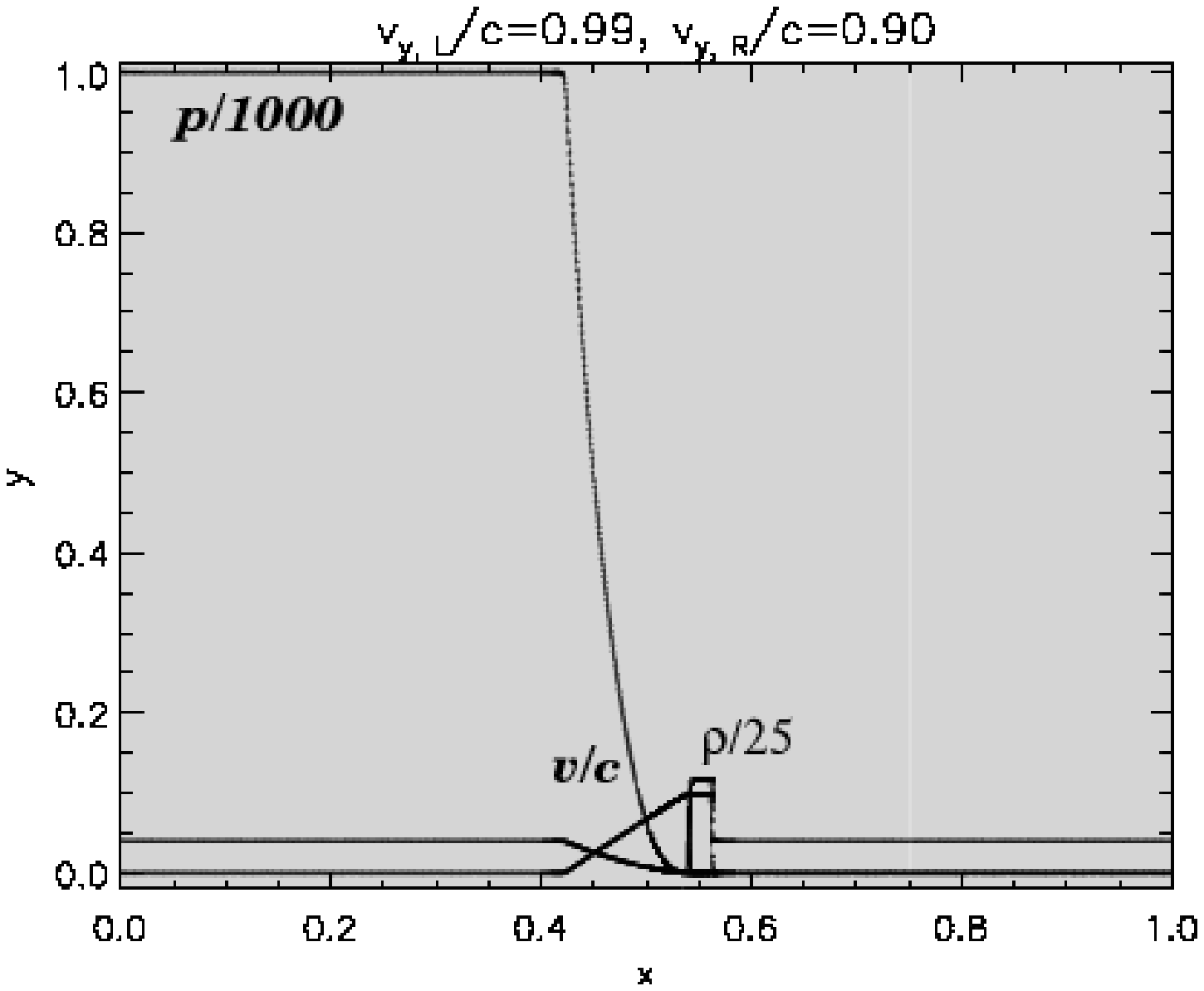}}}}
{\rotatebox{0}{\resizebox{5cm}{3cm}{\includegraphics{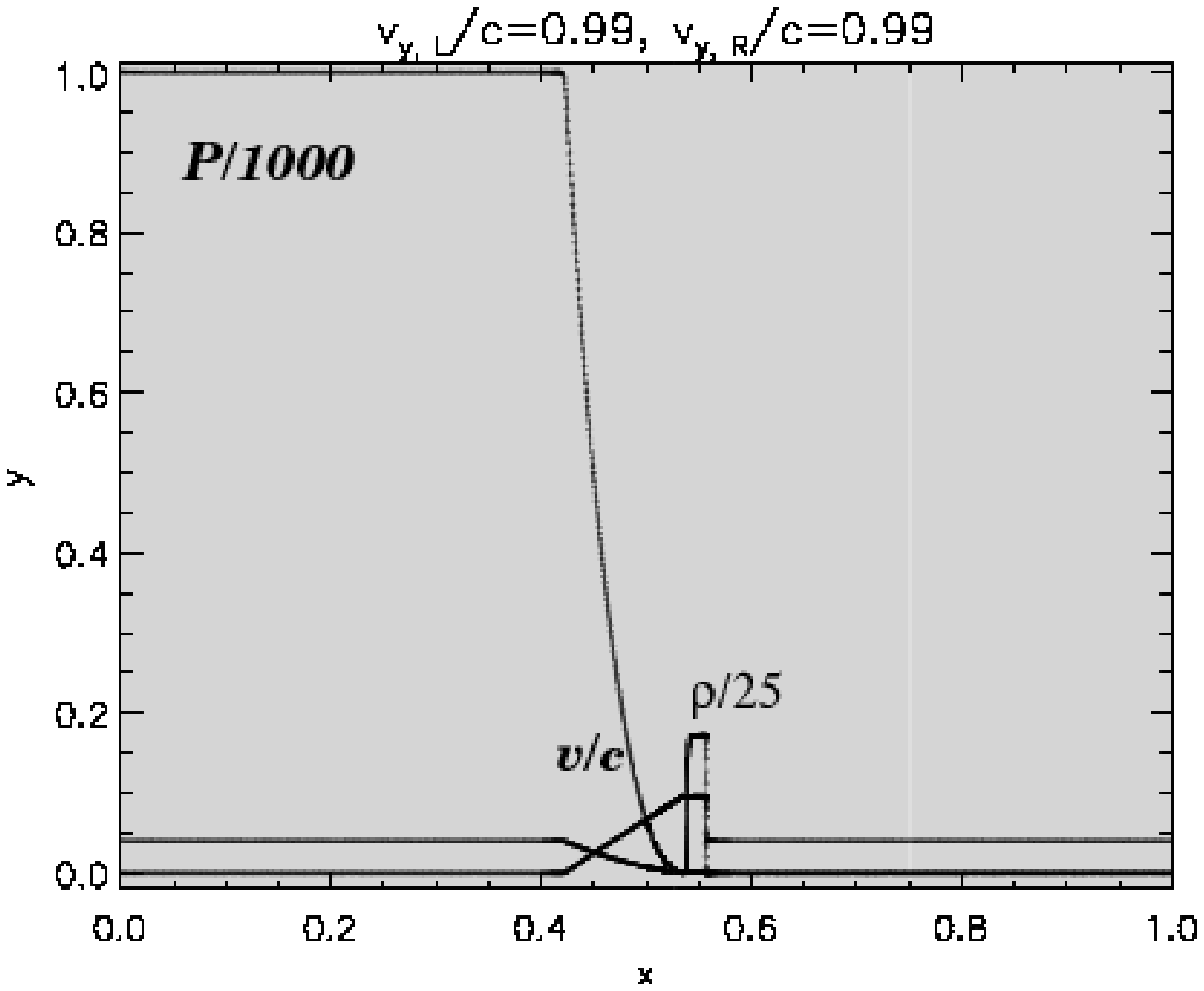}}}}
}
\caption{One-dimensional relativistic shock problems in planar
geometry  with tangential velocities $v_y$. The results presented
 correspond to $t = 0.4$. The solid lines are the analytical solution
\citep{Ponsetal00}; from left to right $v_{\rm y, R} = (0, 0.9,
0.99)c$, and from top to bottom $v_{\rm y, L} = (0, 0.9, 0.99)c$.}\label{LabelFig_2}
\end{center}
\end{figure*}

In a second test, we look particularly into effects due to
nonvanishing tangential velocities, for two ideal gases with
polytropic index $\Gamma = 5/3$. We separate two different constant
states $p_{\rm L} = 10^3$, $\rho_{\rm L} = 1.0$ (left) and $p_{\rm
R} = 10^{-2}$, $\rho_{\rm R} = 1.0$ (right). For the transverse
velocity we form nine combinations  of the pair $v_{y, \rm L}$ and
$v_{y, \rm R}$. As in \citet{Ponsetal00, Mignoneetal05}, we take
$v_{y, \rm L} = (0.0, 0.9, 0.99)c$ in combination with $v_{y, \rm R} = (0.0, 0.9, 0.99)c$.
The spatial separation between the two states is initially at $x = 0.5$. The results at
$t=0.4$ are shown in Fig.~ \ref{LabelFig_2}, where we also overplot the exact solution using the code in \citet{Marti&Muller2003}.

The relativistic effects
in these tests are mainly thermodynamical in the first mild test,
and are due to coupling between the thermodynamics (through specific
enthalpy) and kinetic properties (by the initial tangential velocities).
For small tangential velocity cases, we use only a resolution of $200$ cells on the base
level and $4$ levels. However, for a high tangential velocity case,
we use high base resolution $400$ with 10 levels to resolve the contact
discontinuity and the tail of the rarefaction wave.
In fact, for a high tangential velocity at left (in the high
pressure state), the effective inertia of the left state increases.
This makes the occurring shock move slower and decreases the
distance between the tail of the rarefaction wave and the contact
discontinuity. As also found in~\citet{Zhang&MacFadyen06}, it remains
a numerical challenge to capture the contact discontinuity properly, which we
only managed here by allowing a very high effective resolution.

\subsubsection{Shock Heating Test}
In another 1D test case, a cold fluid hits a wall and a shock front propagates
back into the fluid, compressing and heating it as the kinetic energy
is converted into internal energy. Behind the
shock, the fluid becomes at rest. This test has an analytical solution
in planar symmetry as considered by~\citet{Blandford&McKee76}, and
the jump conditions are
\begin{eqnarray}
p_{2} & = & \rho_{1}  \left(\gamma_{1} -  1\right)
\left(\gamma_{1} \Gamma + 1\right)\,,\nonumber \\
\rho_{2} & =& \rho_{1}  \frac{\gamma_{1} \Gamma + 1}{\Gamma - 1}\,,\nonumber \\
v_{\rm sh} & = &\left(\Gamma - 1\right)\frac{\gamma_{1} v_{1}}{\gamma_{1} +  1}\,.
\label{BM1}
\end{eqnarray}
These give the post shock pressure $p_2$ and density $\rho_2$ values in terms of the incoming density and Lorentz factor, together
with the shock propagation velocity $v_{\rm sh}$.

\begin{figure}
\begin{center}
\FIG{
{\centerline{\rotatebox{0}{\resizebox{8cm}{5cm}{\includegraphics{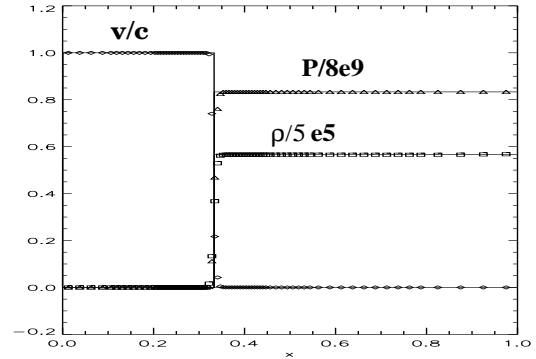}}}}}
} \caption{One-dimensional shock heating problem in planar geometry,
where a cold fluid hits a wall located at $x=1$. The results presented
correspond to $t = 2$. The computational grid consists of $20$ zones with 4
levels of refinement. The solid lines are the analytical solution.}\label{LabelFig_3}
\end{center}
\end{figure}

In our test we take the same initial conditions  as in the recent
paper by~\citet{Zhang&MacFadyen06}, where a cold fluid $p=10^{-4}$
with a density $\rho = 1.0$ has an impact velocity of $v_1 =
\left(1.0 - 10^{-10}\right)$. This corresponds to a Lorentz factor
$\gamma =70710.675$. The temperature after the shock becomes
relativistic, and therefore we take the polytropic index $\Gamma =
4/3$. Hence the shock velocity is $v_{\rm sh} = 0.33332862$. The AMR
simulation is done with $20$ cells on the base level and $4$ levels
on the spatial range $0<x<1$. The result at $t= 2$, with the
reflective wall at $x=1$, is shown in Fig.~\ref{LabelFig_3}. The exact
solution is overplotted as well. In this test, because of the
constant state behind the shock, the maximum impact Lorentz factor
that can be achieved numerically is limited only by the precision of
the Newton-Raphson subroutine. This test is important to demonstrate
its accurate treatment, in view of the intended simulations aimed at
afterglows in GRBs. Indeed, in the shell-frame, the circumburst
medium hits the dense shell with a high Lorentz factor. In a process
similar to what is found in the above test, the kinetic energy of
the impacting medium is converted to thermal energy of
the external medium. Viewed in the lab frame, the swept up
circumburst medium will have similarly high
Lorentz factor and will form a hot shocked layer ahead of the
contact interface. Note also that Fig.~\ref{LabelFig_3} indicates that our discretization
and wall treatment does not suffer from the visible density errors seen in
\citet{Zhang&MacFadyen06}.

\subsection{Two-dimensional tests}
\subsubsection{A relativistic 2D Riemann problem}
A two dimensional square region is divided into  four equal areas
with a constant state each. We fix the polytropic index $\Gamma
=5/3$ and assume free outflow boundary conditions. The relativistic
version of this test was proposed by \citet{DelZanna&Bucciantini02}
and subsequently reproduced by~\citet{Lucas-Serranoetal04,
Zhang&MacFadyen06} and under slightly improved initial conditions by
\citet{Mignoneetal05}. We repeat this simulation with the same
initial configuration from \citet{DelZanna&Bucciantini02}, namely
\begin{eqnarray}
\left(\rho,v_{x}/c,v_{y}/c,p\right)^{\rm NE} &=& (0.1,0.0,0.0,0.01)\,,\nonumber\\
\left(\rho,v_{x}/c,v_{y}/c,p\right)^{\rm NW} & =& (0.1,0.99,0.0,1.0)\,,\nonumber\\
\left(\rho,v_{x}/c,v_{y}/c,p\right)^{\rm SW} &=& (0.5,0.0,0.0,1.0)\,,\nonumber\\
\left(\rho,v_{x}/c,v_{y}/c,p\right)^{\rm SE} &=& (0.1,0.0,0.99,1.0)\,.
\end{eqnarray}

\begin{figure}
\begin{center}
\FIG{
{\rotatebox{0}{\resizebox{8cm}{7cm}{\includegraphics{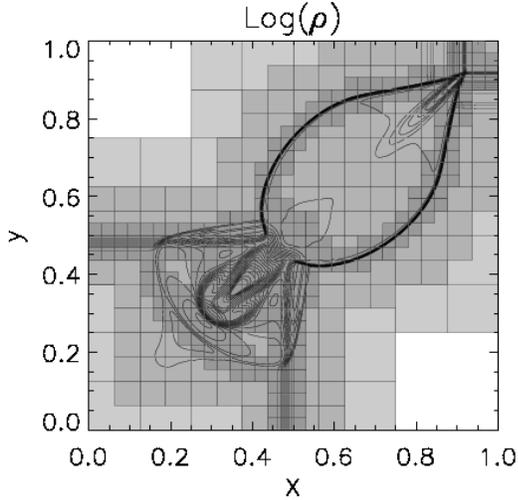}}}}
} \caption{Density distribution for the two dimensional shock tube
problem at $t=0.4$. With a polytropic index $\Gamma= 5/3$, a base
resolution of $48\times 48$ and $4$ AMR levels.}\label{LabelFig_4}
\end{center}
\end{figure}

The simulation is done with $48\times48$ cells at the lowest grid
level, and we allow for $4$ levels. The result is shown in
Fig.~\ref{LabelFig_4}. Our result is in qualitative agreement with
those results published, and shows the stationary contact
discontinuities between SW-SE and SW-NW with a jump in the
transverse velocity. These are somewhat diffused by the employed
{ Total Variation Diminishing Lax-Friedrichs (TVDLF) discretization}
(T\'oth \& Odstr\v cil 2006). A simple and easily affordable remedy
for improvement is to activate many more grid levels. Shocks feature
across the interfaces NW-NE and SE-NE, propagating diagonally to the
NE region, and an elongated diagonal shock structure forms as the NE
sector recedes into the RHS top corner. In the SW corner, an oblique
jet-like structure forms with a bow shock.

\subsubsection{Relativistic jet in 2D cylindrical geometry}
\begin{figure}
\begin{center}
\FIG{
{\rotatebox{0}{\resizebox{8cm}{14cm}{\includegraphics{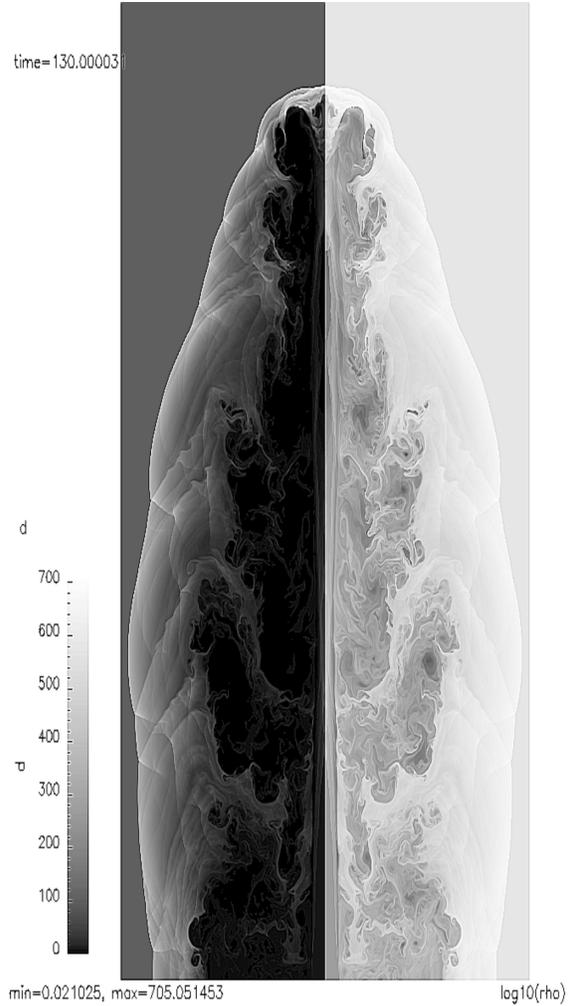}}}} }
\caption{Density distribution for the axisymmetric relativistic jet
at $t=130$. At left, we show the lab frame density, at right, we
show the proper density in a logarithmic scale. The computational
base grid consists of $90 \times 300$ zones with 5 levels of
refinement and the domain size is $15 \times 50$.} \label{LabelFig_5}
\end{center}
\end{figure}

Since it is relevant for our 2D GRB simulations, we also present a
two-dimensional simulation of an axisymmetric relativistic jet
propagating in a uniform medium. We simulate the C2 jet model
from~\citet{Martietal97}, but with an enlarged domain and at higher
effective resolution.
Our computational domain covers the
region $0<r<15$ and $0<z<50$ jet radii. Initially, the relativistic
jet beam occupies the region $r \leq 1, z\leq 1$, with
 $v_{\rm jet} = 0.99$, $\rho_{\rm jet} = 0.01$ and its classical
Mach number $M = 6$. In this case, the jet is super-sonic but its
temperature is still classical, so we can take the polytropic index
$\Gamma = 5/3$. The density of the external medium is $\rho_{\rm ext}=1.0$. We follow the evolution until $t=130$, and
this end result is shown in Fig.~\ref{LabelFig_5}.
We performed the simulation with a resolution at the lowest
level of the grid set to $90 \times 300$, and allowed for a total of
5 levels of refinement eventually achieving an effective
resolution of $1440 \times 4800$.

 In this simulation, the relativistic motion of the flow dominates,
the thermal energy is weak compared to the kinetic energy. As a
result the external medium influences only weakly the jet and the Lorentz
factor $\gamma\sim 7$ flow produces a cocoon
structure from the tip of the jet. One also finds a weak transverse
expansion of the outflow in accord with what is reported
by~\citet{Martietal97}. This transverse expansion of the jet is
induced by the pressure build-up inside the cocoon
\citet{Begelman&Cioffi89}. In our simulation, the average transverse
expansion obtained is occurring at an estimated speed of
$v_{\rm T}=0.11\, c$. Moreover, at the contact interface between shocked
 external medium and jet material, complex vortical structures form.
These originate from Kelvin-Helmoltz type instabilities, as a
consequence of cold fast jet outflow meeting a more static medium.
The average propagation speed of the jet head is found to be
$0.414\,c$, which is in agreement with the one-dimensional
analytical estimate of $0.42\,c$ as given by~\citet{Martietal97}.

\begin{figure}
\begin{center}
\FIG{
 {\rotatebox{0}{\resizebox{9cm}{5cm}{\includegraphics{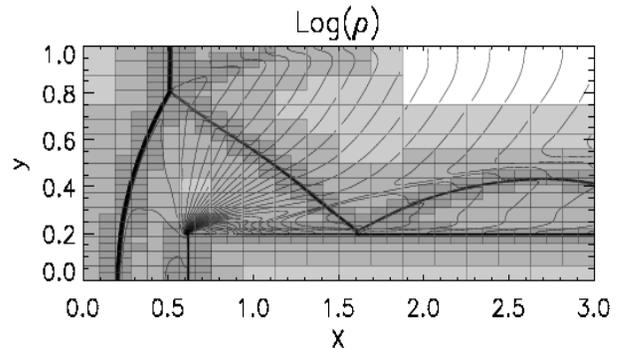}}}}
} \caption{Density distribution, in logarithmic scale, for the
forward facing step problem, at $t=4.26$.}\label{LabelFig_6}
\end{center}
\end{figure}
\subsubsection{Wind tunnel with step}
We reproduce here a standard test in the hydrodynamic literature, namely the
forward facing step test
from~\citet{Emery1968, Woodward&Colella1984}, but adjusted to the
relativistic hydro regime as in~\citet{Lucas-Serranoetal04, Zhang&MacFadyen06}.
 A horizontal relativistic supersonic flow enters a tunnel with
a flat forward facing step. The test was done with a resolution $50
\times 100$ zones with $4$ levels. The size of the tunnel is $0 \leq
x \leq 3$ and $0 \leq y \leq 1$. The step is $0.2$ in height and its
position is at $x=0.6$. It is treated as a reflecting boundary. The
upper boundary and lower boundary for $x < 0.6$ are also both
reflecting. However, the left boundary is fixed at the given inflow
and the right one has free outflow. Initially, the whole
computational domain is filled with ideal gas with $\Gamma=7/5$ with
a density $\rho=1.4$ moving at $v_{x}=0.999$,  i.e., with a Lorentz factor
 $\gamma=22.37$. The Newtonian Mach number is set to $3$. The result of our
simulation is shown in Fig.~\ref{LabelFig_6} at time $t=4.26$.
In this test, the relativistic flow collides with the step, as a
result a reverse shock propagates back against the flow direction and this
shock reflects from the upper boundary. A Mach stem forms and remains
stationary.
The result of our simulation is comparable to what is reported in
\citet{Lucas-Serranoetal04}.

\begin{figure*}
\begin{center}
\FIG{
{\rotatebox{0}{\resizebox{5.7cm}{5cm}{\includegraphics{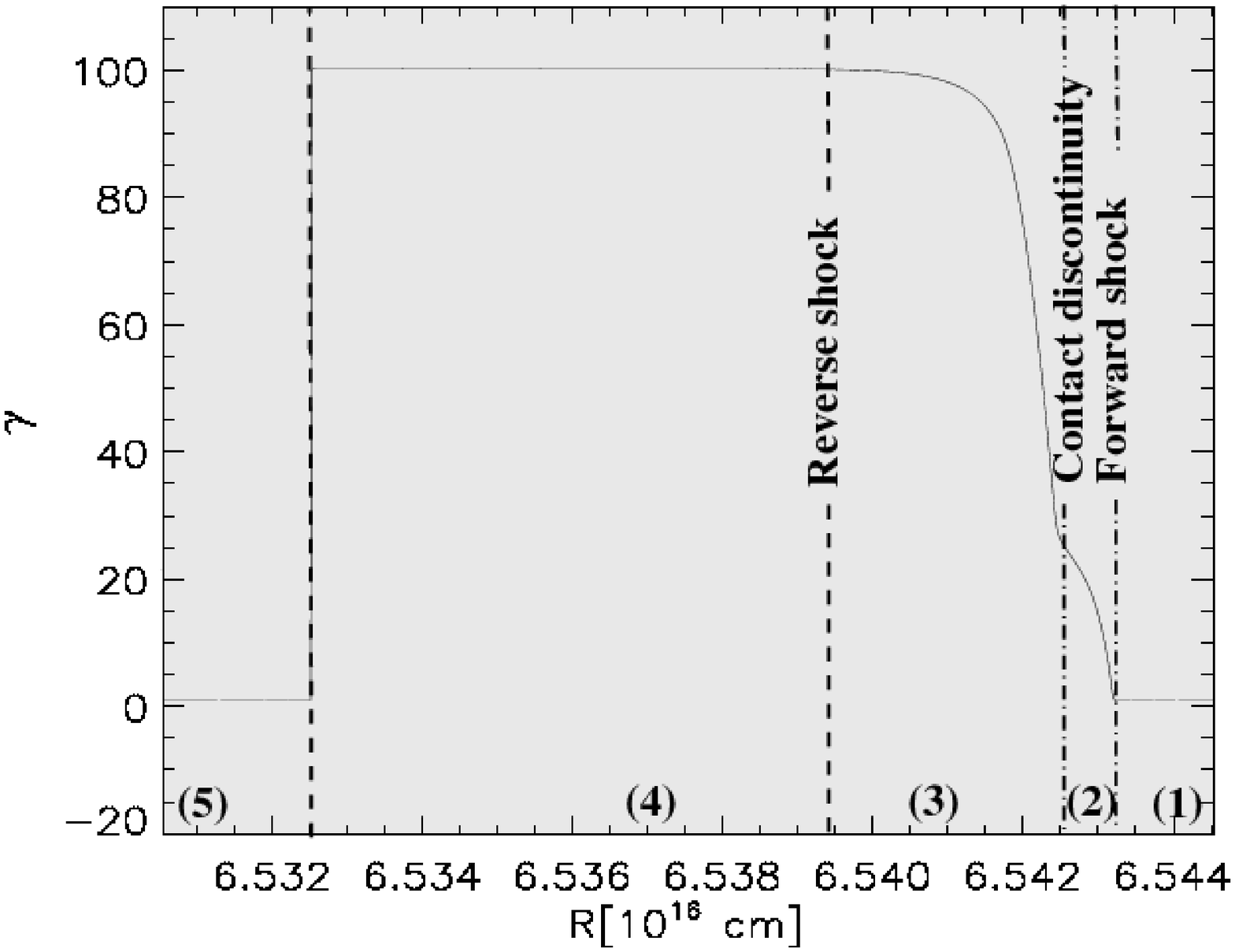}}}}
{\rotatebox{0}{\resizebox{5.7cm}{5cm}{\includegraphics{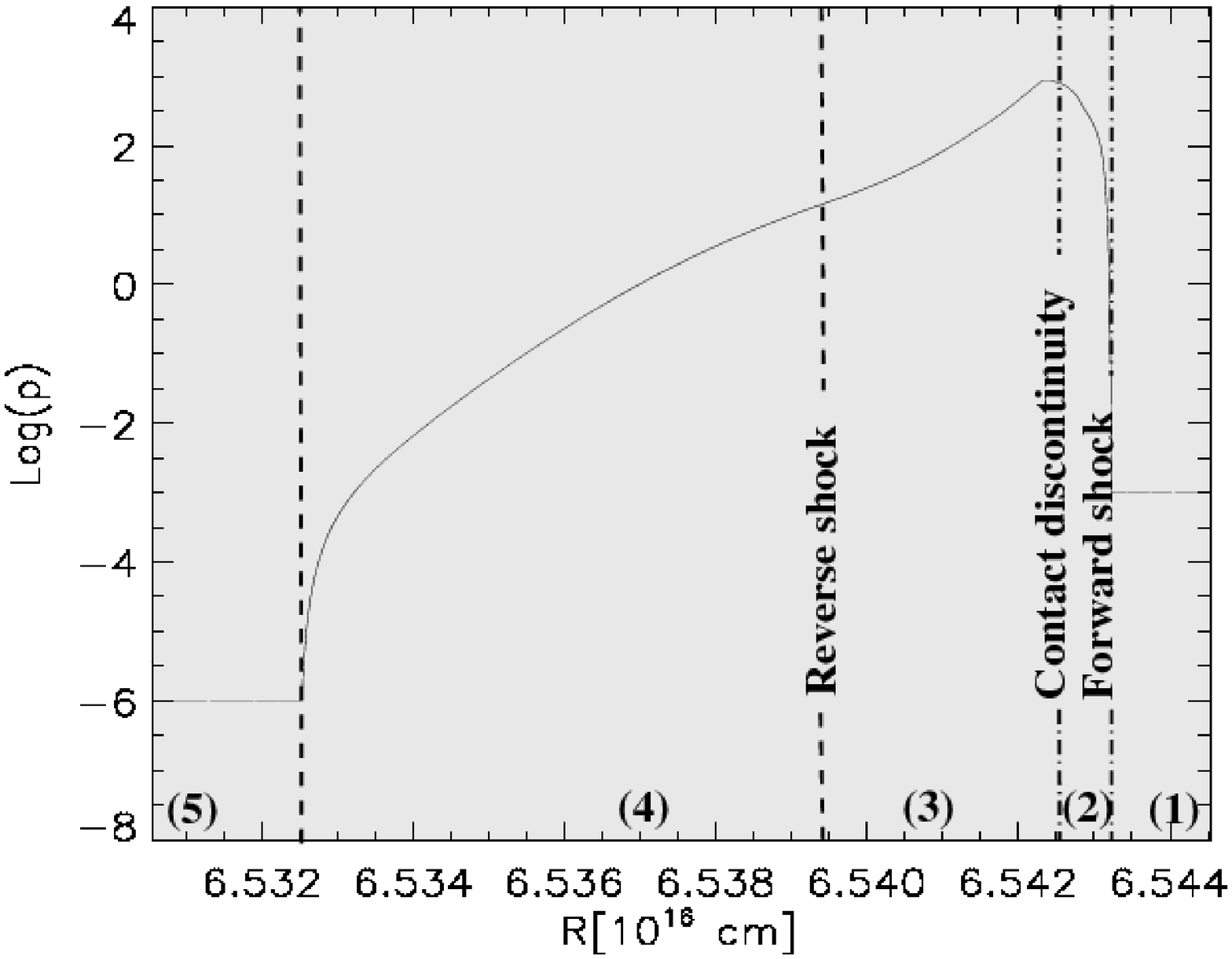}}}}
{\rotatebox{0}{\resizebox{5.7cm}{5cm}{\includegraphics{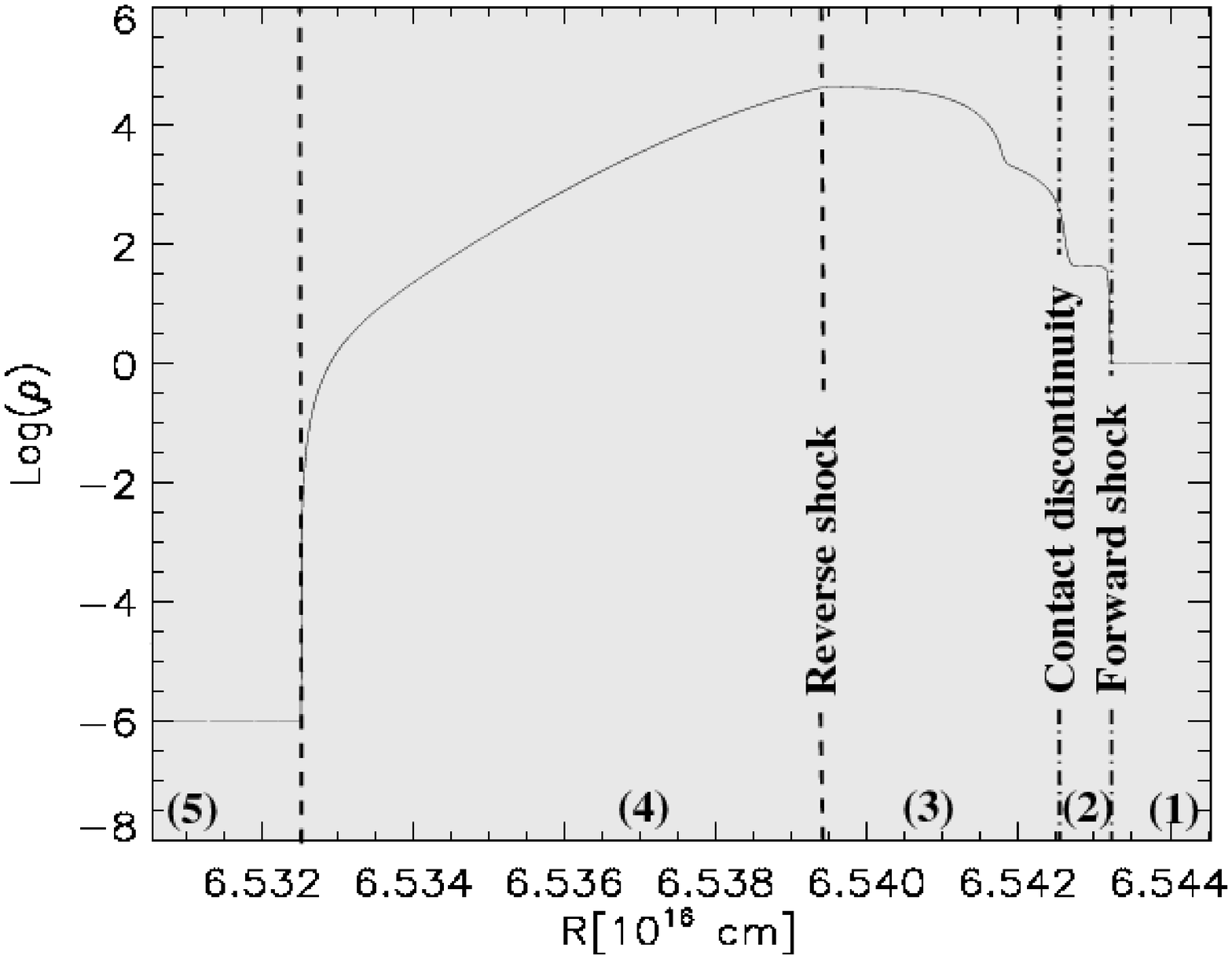}}}} }
\caption{The five regions characterizing the interaction of the
relativistic shell  with the ISM at $t=2.2\times 10^{6} {\rm s}$.
Panel(a): Lorentz factor, (b): logarithm of the pressure, (c):
logarithm of density.}
\label{LabelFig_7}
\end{center}
\end{figure*}
 \begin{figure*}
\begin{center}
\FIG{
{\rotatebox{0}{\resizebox{6cm}{4cm}{\includegraphics{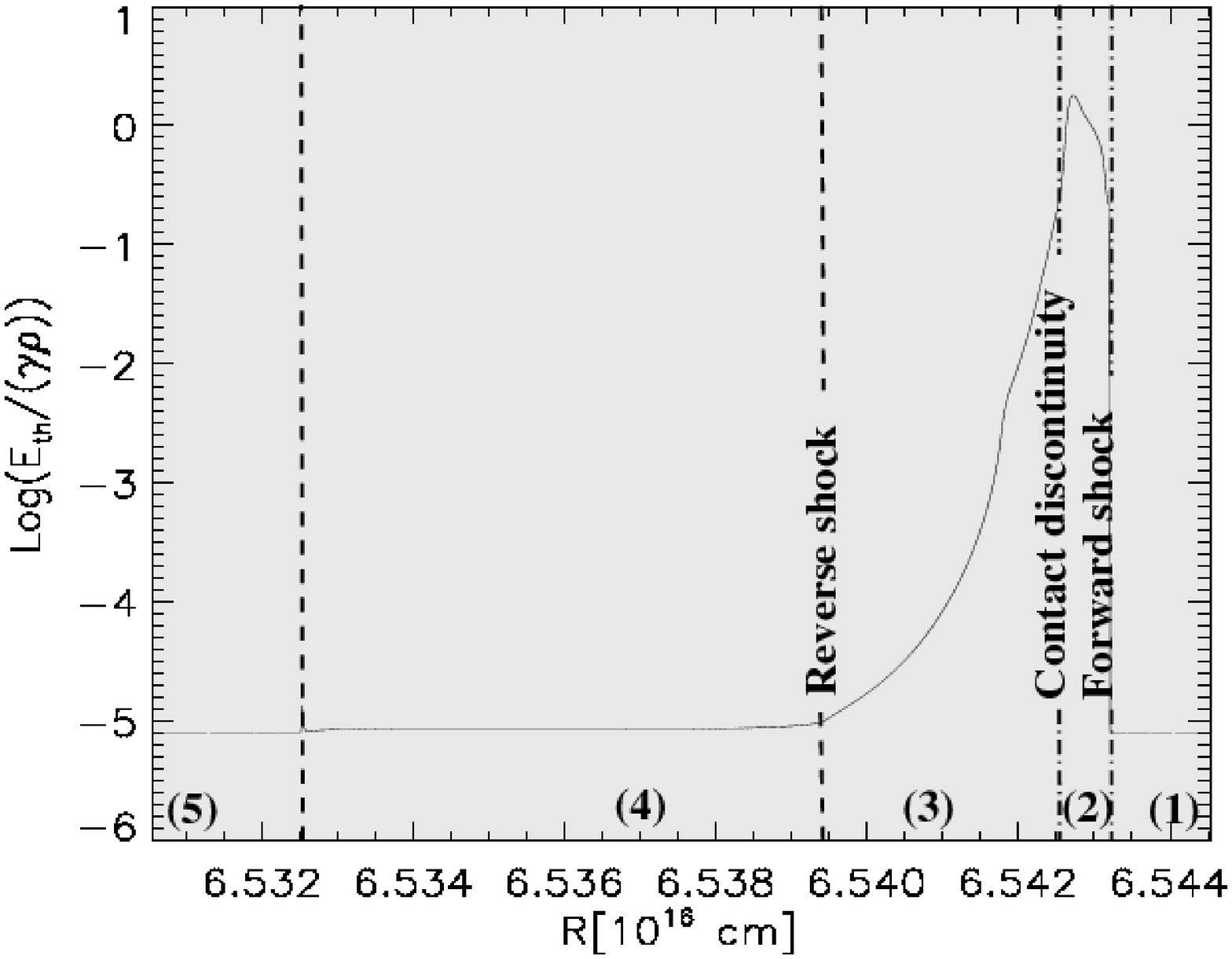}}}}
{\rotatebox{0}{\resizebox{6cm}{4cm}{\includegraphics{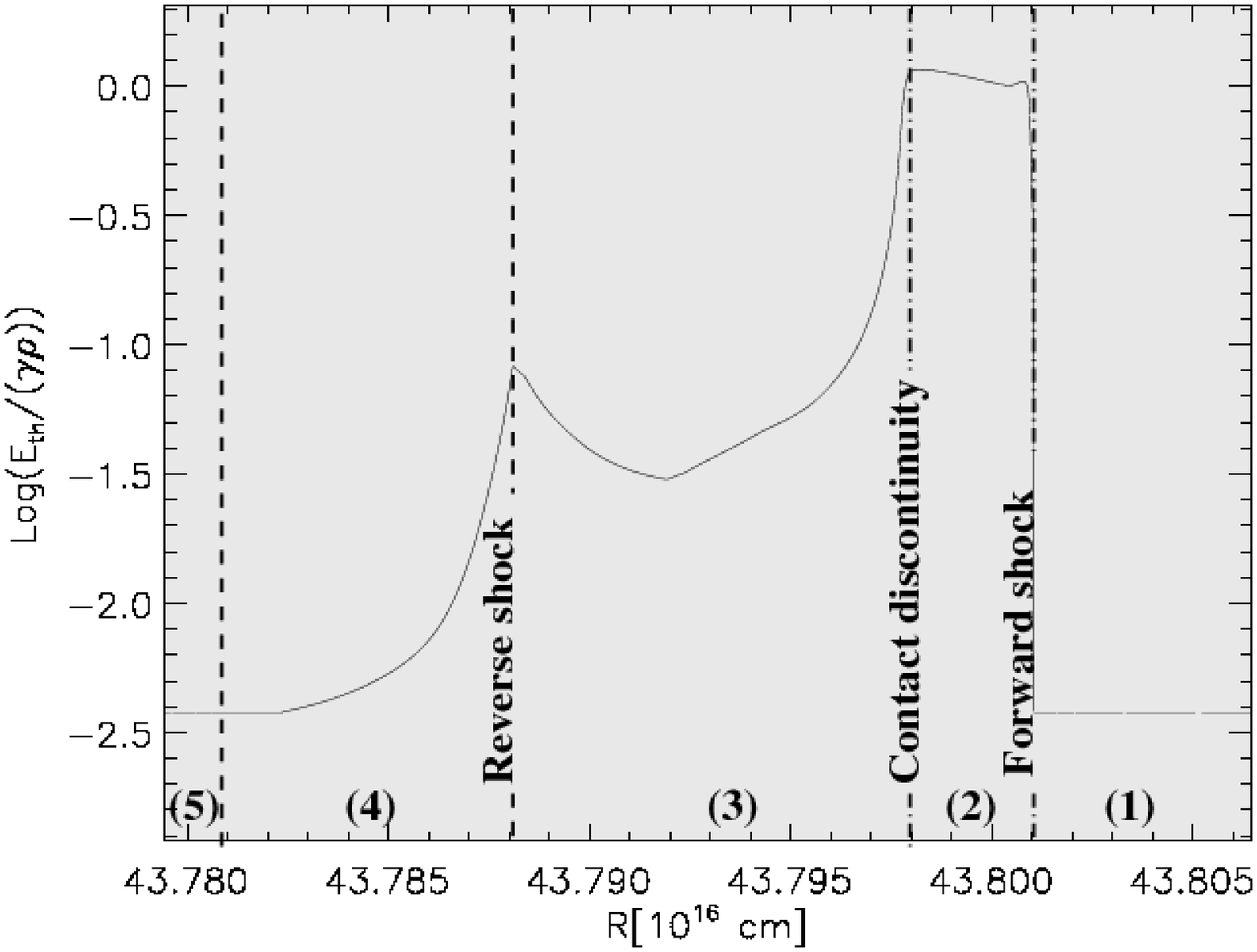}}}}
}
\caption{ Ratio of  thermal energy to mass energy, when
the shell reaches distance: $R\simeq 6.5\times 10^{16}{\rm cm}$
($t\simeq 2.2\times 10^{6} {\rm s}$) (left), and $R\simeq 4.3\times
10^{17}{\rm cm}$ ($t\simeq 1.5\times 10^{7} {\rm s}$) (right).}
\label{LabelFig_8}
\end{center}
\end{figure*}
\section{GRBs and models for their afterglow phase}\label{after}
 A popular model for GRB flows is known as the fireball
model. In this model, GRBs are produced by a relativistic outflow
following a violent event near a compact object. A large amount of
energy is promptly released by the compact source in a region with
small baryon loading (for a review see \citet{Piran05}).  Initially,
most of the energy of the flow is in the form of internal (thermal) energy. The shell
expands rapidly converting its internal energy to kinetic. After the
acceleration phase is complete, the shell is cold and moves with relativistic
speeds.

This cold shell interacts with the circumburst medium,
producing strong shocks. Our simulations will consider the dynamics
from this phase onwards. As the shell sweeps up mass from the external medium,
the kinetic energy in the relativistic shell is gradually transferred to
kinetic and internal energy in the shocked ambient medium. Moreover,
the shell itself gets traversed by a reverse shock, which
in turn converts the kinetic energy of the shell to internal energy.

The observed afterglow emission that follows the prompt GRB emission
is believed to come from synchrotron (with possible inverse Compton
contribution) emitting electrons that are accelerated in the forward
and reverse shocks \citep{Sarietal98, Galamaetal98}. In the initial
phases of the shell-ISM interaction, the electrons can be in the
fast cooling regime (i.e. their cooling timescale is shorter than
the expansion timescale) and, therefore, radiate efficiently most of
the energy injected to them. Furthermore, if most of the energy
dissipated in the shocks accelerates the electrons, then one has to
consider radiative shocks. If either of the previous conditions does
not hold, the radiative losses in the shocks are small. Here, we
assume that the radiative losses are dynamically unimportant,
 i,e., the shocks are adiabatic throughout these simulations.
{ According to the magnetization of the shell, the interaction
shell-ISM and the spectrum could change as is shown in
\citet{Mimicaetal06}. Here, we assume that the magnetic field is
dynamically unimportant.}
\subsection{1D isotropic shell evolution}
\begin{figure*}
\begin{center}
\FIG{
{\rotatebox{0}{\resizebox{5.7cm}{5cm}{\includegraphics{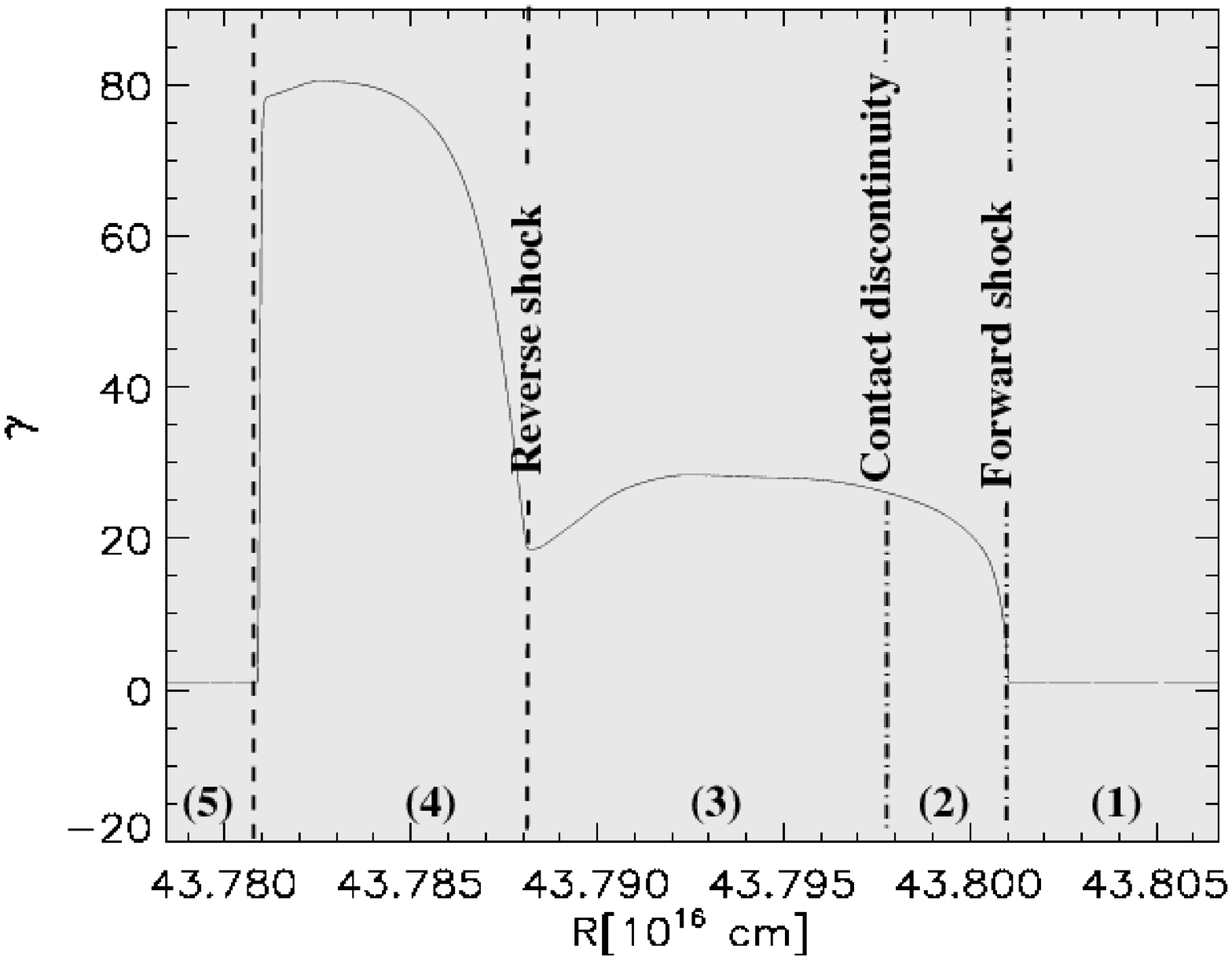}}}}
{\rotatebox{0}{\resizebox{5.7cm}{5cm}{\includegraphics{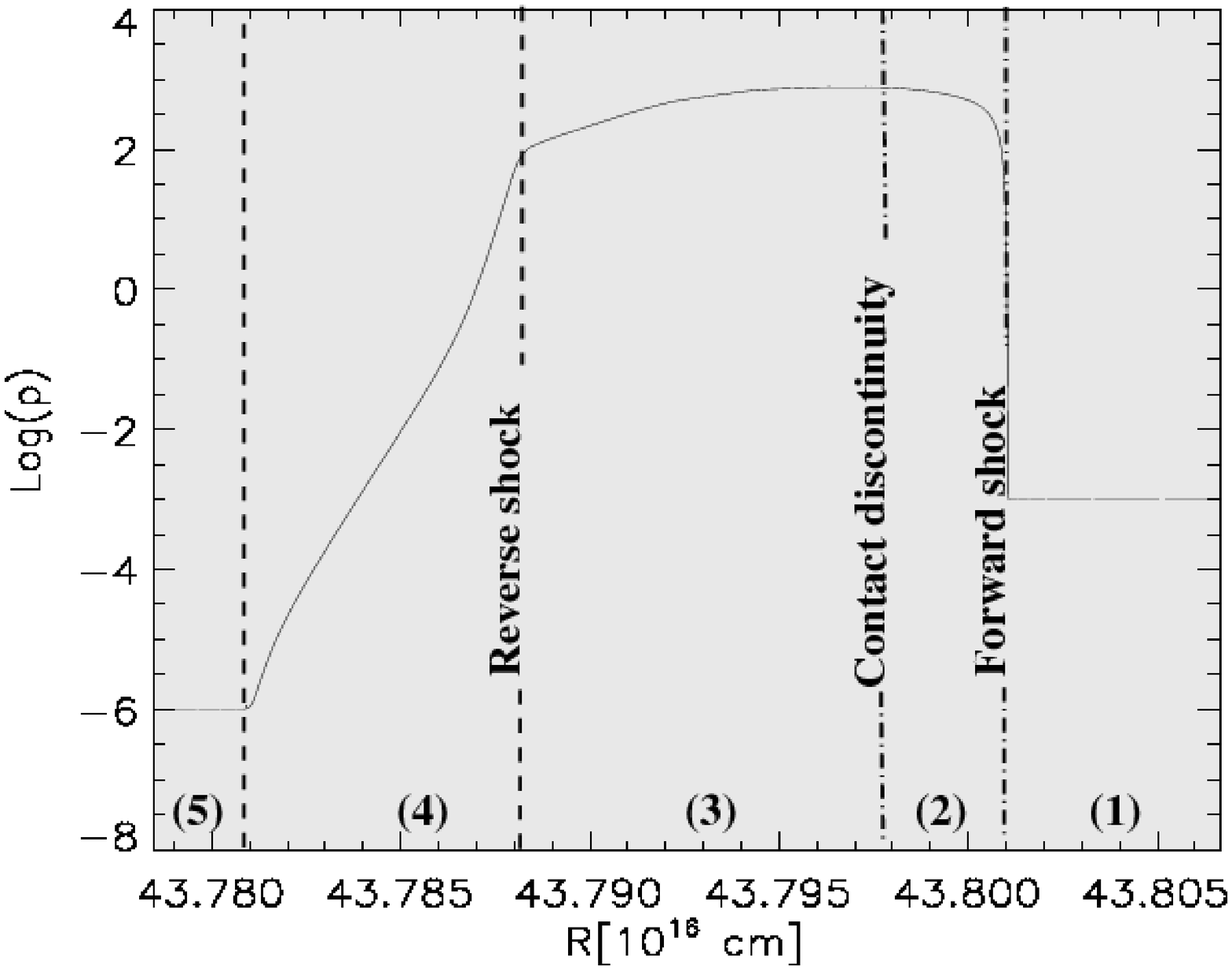}}}}
{\rotatebox{0}{\resizebox{5.7cm}{5cm}{\includegraphics{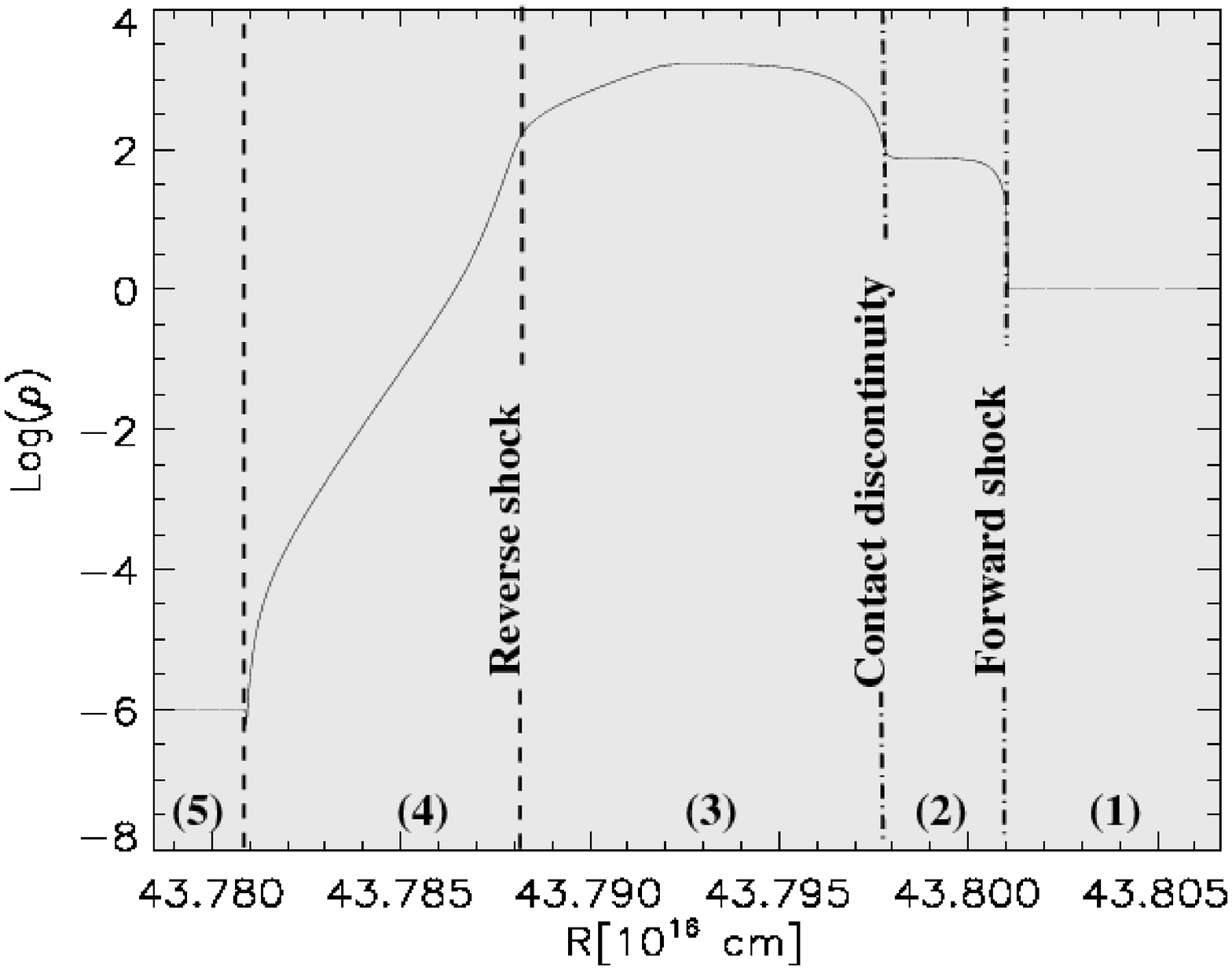}}}}}
\caption{The five zones present when the relativistic shell
interacts with the ISM at $t\simeq1.5\times 10^{7} {\rm s}$. (a)
Lorentz factor, (b) log of pressure, (c) log of density.}
\label{LabelFig_9}
\end{center}
\end{figure*}

In this simulation, we consider an ISM with uniform number density $
n_{\rm ISM}=1\, {\rm cm}^{-3}$. Many GRB afterglows (more than $25\%$)
seem to be produced in such constant density medium
\citep{Chevalier&Li00, Panaitescu&Kumar02, Chevalieretal04}. This
constant density medium can be the resultant of a Wolf-Rayet star
progenitor, with its surroundings shaped by a weak stellar wind
\citep{VanMarleetal06}. Initially we set a uniform relativistic shell
 at $R_0=10^{16}{\rm cm}$ from the central engine, since according
 to \citet{Woods&Loeb95}
the interaction of the shell with the ISM becomes appreciable at
this distance. The shell has an initial Lorentz factor of
$\gamma=100$  (a $\gamma\ge 100$ is in accord with a shell which is
optically thin to  gamma-rays \citep{Woods&Loeb95, Sari&Piran95}),
and energy
\begin{equation}
E=10^{54}{\rm ergs}=4\pi \gamma^2\, R_0^2 \delta \, \rho_{\rm shell}
c^2,
\label{Eiso}
\end{equation}
where $\delta$ stands for the lab-frame thickness of the shell set to
$5 \times 10^{12}{\rm cm}$ is of the order of the expected value for a fireball
 $\delta\sim max(c\,\Delta t, R_0/\gamma^2)$, where $\Delta t$ is the duration of the GRB.
 The ISM and the shell are cold, and the initial pressure is set to
$p_{\rm ISM}=10^{-3} n_{\rm ISM}\,m_{p}\,c^2$ and $p_{\rm shell}=10^{-3}
n_{\rm shell}\,m_{p}\,c^2$ respectively.
Note that this implies a huge initial contrast in the density measured
in the lab-frame between the shell and the ISM $D_{\rm shell}/D_{\rm
ISM}\sim 10^9$, and this presents an extreme challenge from a computational
point of view.
 Initially, the energy of the shell is then mainly kinetic. We
use a constant polytropic index $\Gamma=4/3$, as the interaction
shell-ISM will be dominated by the forward shock, where the
temperature of the shocked ISM becomes relativistic.

 In this simulation we use an effective resolution of 1536000 cells
corresponding to the highest grid level 10 allowed. We use the full
AMR capabilities in this simulation, since we simulate on a domain
of size $[0.3,300] \times 10^{16}{\rm cm}$, with 30000 grid points
on the lowest level. At $t=0$, the shell itself is then only
resolved from grid level 6 onwards, when we use a refinement ratio
of 2 between consecutive levels.
The initial shell is resolved by about 25 cells in grid level 10
(later in the dynamical evolution this means that there are many more grid
points throughout the widening structure).
We use this very high effective resolution to avoid any numerical diffusion
which may cause an artificial spreading of the shell.
We ensure that throughout the entire simulation, grid level $10$ is activated
and concentrates fine grids on both the forward shock and reverse shock
regions. Both are very important to determine the precise timing of the
deceleration.

In Fig.~\ref{LabelFig_7}, we show snapshots taken at lab-frame time
$t \simeq 2.2\times 10^{6} {\rm s}$ corresponding to an early time in the entire
simulation, and in Fig.~\ref{LabelFig_9}, we show snapshots taken at time
$t\simeq 1.5\times 10^{7} {\rm s}$ corresponding to a time when the shock is fully
developed, we will concentrate our discussion mainly on this
figure~\ref{LabelFig_9}.
These figures demonstrate that we resolve all four regions
that characterise the interaction between an outward moving
relativistic shell and the cold ISM. From right to left, we
recognize (Fig.~\ref{LabelFig_9}) (1) the ISM at rest, (2) the shocked ISM that has passed
through the forward shock, with its Lorentz factor raised to
$\gamma_{(2)}\sim 30$. This swept-up ISM gets compressed at the
front shock and its number density reaches $ n_{(2)}\sim 75{\rm cm^{-3}}
\sim \frac{\Gamma \,\gamma_{(2)}+1}{\Gamma-1}$ \citep{Sari&Piran95}.
These two values correspond to the analytical estimate given by
eq.~(\ref{BM1}) for the front shock propagation. Region (3) represents
part of the initial shell material which is shocked by the reverse
shock.  The reverse shock propagates back into the cold shell,
reducing its Lorentz factor and converting its kinetic to thermal
energy. Transfer of energy from the initial cold shell thus occurs
both at the forward and the reverse shock. Regions (2) and (3) are
separated by a contact discontinuity \citep{Meszaros&Rees92}. At this
spherical contact surface, the longitudinal velocity (Lorentz factor
in 1D case) and pressure remain constant, but there is a jump in
density. Furthermore, region (4) is the unshocked cold material  of the shell,
moving with a Lorentz factor $\gamma_{(4)}=100$. The weak thermal
energy of the shell interior itself did induce a slight expansion in
the thickness of this part of the shell.

\begin{figure}
\begin{center}
\FIG{
{\rotatebox{0}{\resizebox{8cm}{6cm}{\includegraphics{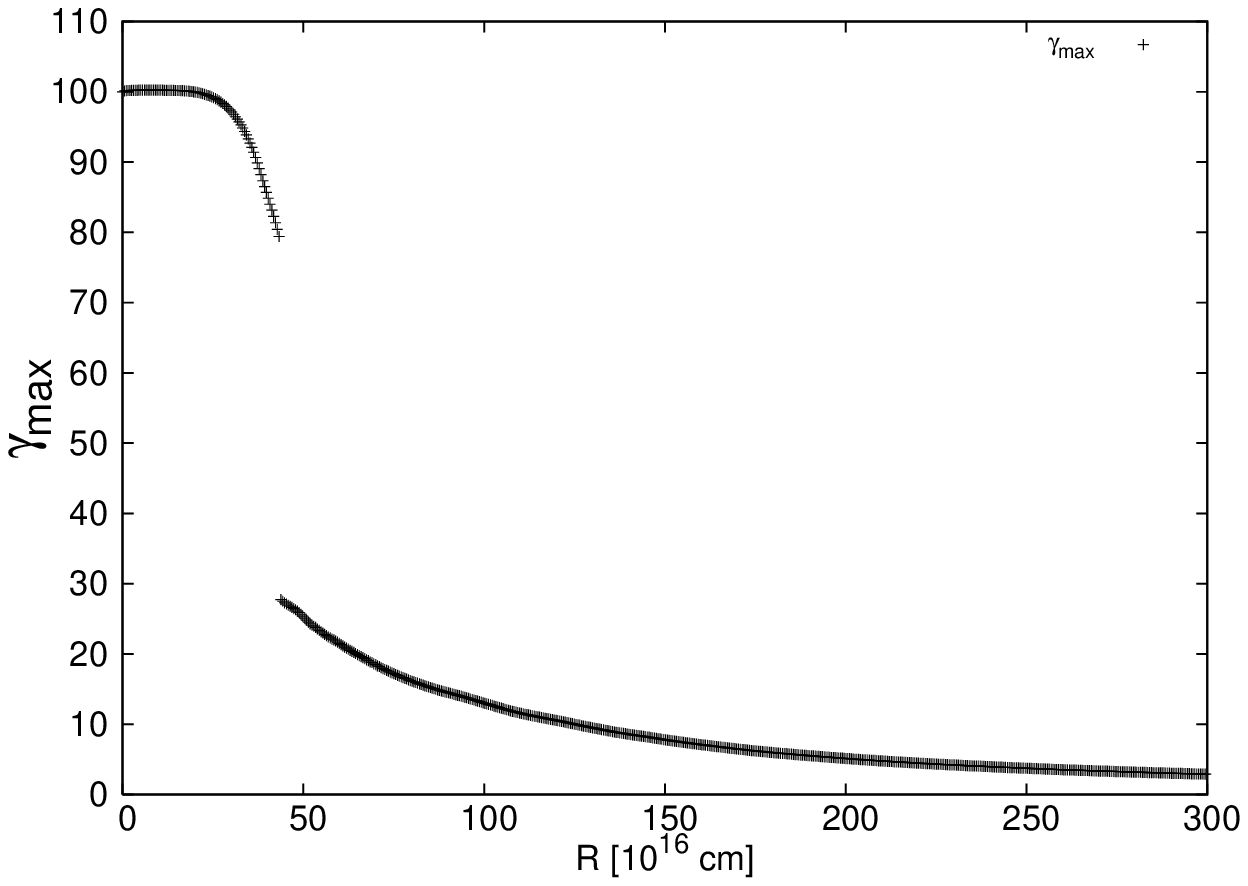}}}}
{\rotatebox{0}{\resizebox{8cm}{6cm}{\includegraphics{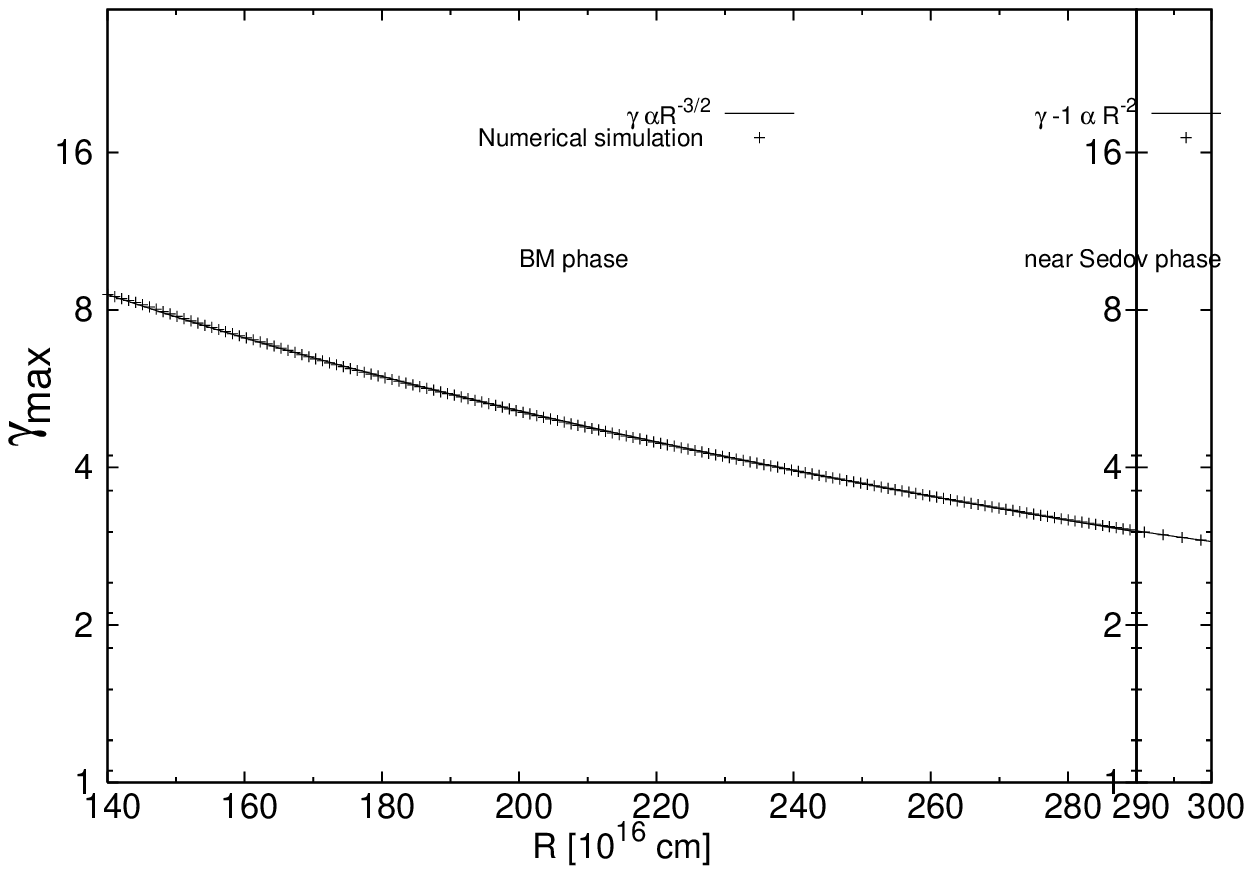}}}} }
\caption{The variation of the maximal Lorentz factor in the
propagating shell-ISM structure with time a) when the shell
propagates from $R_{0}=10^{16}{\rm cm}$ to $R=300 \times R_{0}$ b)
 when the shell decelerates following the  Blandford-McKee profile.}
 \label{LabelFig_10}
\end{center}
\end{figure}
The reverse shock
separating region (3) and (4) propagates into the cold shell with a
 Lorentz factor $\gamma_{RS}\,= \,\gamma_{(3)}\,\gamma_{(4)}\,
\left(1-v_{(3)}v_{(4)}/c^2\right)\sim\,2.5$. This reverse shock
is Newtonian inefficient in raising the thermal energy content
as is shown in Fig.~\ref{LabelFig_8} (left panel), where
we draw the specific thermal energy in the shocked ISM and shell
when the shell reaches a  distance $R\simeq 6.5 \times 10^{16}{\rm cm}$.
The reverse shock remains Newtonian
until it reaches a distance from the GRB source of $R\sim 3.8\times
10^{17}{\rm cm} $. Then it becomes mildly relativistic until $R\sim
4.3\times 10^{17}{\rm cm} $ where the reverse shock becomes very
efficient to convert the kinetic energy to thermal energy
(see Fig.\ref{LabelFig_8} at right). Beyond this latter
distance, the density of the unshocked shell part $\rho_{(4)}$ has
decreased in accord with the spherical expansion of the shell,  to
$\rho_{(4)}\ll \gamma_{(4)}^2\, \rho_{\rm ISM}$. As a result, the
reverse shock becomes relativistic. This behavior is characteristic
for an initial thin cold relativistic shell  decelerating in a constant
density external medium. In fact, until the
outward propagating shell reaches $R\sim 3.8\times 10^{17}{\rm cm}
$, the shocked ISM matter is hot $e_{(2)} \sim 30.0 \rho_{(2)}$
(where $\gamma_{(2)}=30$ corresponds to the analytical solution for
the relativistic forward shock $e_{(2)}=\left(\gamma_{(2)}-1\right)
\rho_{(2)}$), while the shocked shell material which has
$e_{(3)}=e_{(2)}$ is cold, since $e_{(3)}\sim 0.01 \rho_{(3)}$. When
the density in the non-shocked shell (i.e. $ \rho_{(4)}$) decreases
enough due to spherical expansion, the Lorentz factor of the reverse
shock increases and the last part of the shocked shell becomes hot
$e_{(3)}\sim \rho_{(3)}$.

 There is another region $(5)$ indicated in the figures behind the shell.
The density and the pressure in the region $(5)$ are very small with
$n_{(5),\rm min}<10^{-6}{\rm cm^{-3}}$ and  $p_{(5),\rm min}<10^{-6}
m_{\rm p}c^{2}$. Therefore, the region $(5)$ is in the numerical
point of view a vacuum. This region $(5)$ is not of strong interest
for the physics of the afterglow, but it is computationally
challenging to resolve the interface between the regions $(4)$ and
$(5)$ where the ratio of the lab frame density between the two
reaches $D_{(4)}/D_{(5)}\sim 10^{14}$
{ in the first phase of the propagation of the shell ($R\sim
10^{16}{\rm cm}$), while the expansion of the shell remains weak. By the time shown in Fig.~\ref{LabelFig_7}, this contrast has dropped to a value of at most $10^{12}$.}

The near-total deceleration of the
shell only takes place when the two shocks in the shell-ISM
interaction manage to convert an important fraction of the kinetic
energy of the shell to thermal energy (and the efficiency of this
conversion depends on whether the reverse shock is relativistic or
Newtonian, as discussed above), while the rest is transferred to the
swept-up ISM in the form of kinetic and thermal energy. In the first
phase of the deceleration, the maximum Lorentz factor of the shell
decreases gently from $100$ at a distance of $R\sim 2.5\times
10^{17}{\rm cm}$ to 80 at a distance of $R\sim 4.3 \times
10^{17}{\rm cm}$.  However, only at the latter distance  of $4.3
\times 10^{17} {\rm cm}$, a sudden decrease of the maximum Lorentz
factor of the entire configuration
from $\gamma=80$ to $\gamma=30$ takes place. This fast drop of the maximum
Lorentz factor as seen in Fig.~\ref{LabelFig_10}
coincides with the moment at which the reverse shock reaches the
back end of the cold shell, thereby converting its kinetic to
thermal energy (Fig.~\ref{LabelFig_9}). In fact, when
plotting the maximal Lorentz factor as a function of distance,
initially we always observe the Lorentz factor of the unshocked
shell matter. As soon as the reverse shock has crossed the entire
initial shell, we start to follow the evolution of the Lorentz
factor of the shocked ISM (at the forward shock)  where the maximum
Lorentz factor is $30$ at that particular moment.

After this phase, the shell structure continues to decelerate by
transferring its kinetic energy to shocked ISM matter at the forward
shock. However, as seen in Fig.~\ref{LabelFig_10}, it
still takes a certain time  before the
variation of maximum Lorentz factor now characterizing the shocked
ISM matter follows the self-similar analytical solution for
blast-wave deceleration as put forward by \citet{Blandford&McKee76}.
From about a distance of $1.2\times
10^{18}{\rm cm}$, our numerical solution starts to follow the
analytical solution precisely. In fact, after the reverse shock
traversed the entire initial shell, a forward traveling rarefaction
wave propagates through the entire structure thereby slowing it down
while transferring most of the energy to shocked ISM regions. This
structure does not follow the self-similar prescription and causes
the initial difference. In the end, the distance between the forward
shock and the contact discontinuity increased sufficiently and the
resulting radial thermodynamic profiles in between become fully
described by the Blandford-McKee analytical solution.  The
Lorentz factor predicted by the Blandford-McKee solution behind the forward
shock (for an adiabatic shock) is
$\gamma_{\rm BM}=(E/ \rho_{\rm ISM}c^2R^3)^{1/2}\propto R^{-3/2}$.
The prediction of the
Blandford-McKee solution for the Lorentz factor of the flow is also plotted in
Fig.~\ref{LabelFig_10} and the agreement with the results of the
simulation at these later stages of the decelation is good.

Eventually, we
enter into the mildly relativistic regime for the blast wave
evolution. The transition to the
Sedov-Taylor phase occurs beyond the simulated distance $R>300\times
10^{16}{\rm cm}$, since we still have a Lorentz factor of about 3 at
the end of the simulation. The Sedov-Taylor distance we
find is close to the analytical estimate given by $l\sim
(3E/ 4\pi \rho_{\rm ISM}\,c^2)^{1/3}\sim 5\times 10^{18}{\rm cm}$.

\subsection{2D modeling of directed ejecta}

Precise analysis of the afterglow phases requires to evolve
numerically confined ejecta in more than 1D, propagating in a
jet-like fashion  into the ISM. We now
present axisymmetric, 2D simulations of a relativistic cold shell
propagating in uniform ISM with a number density $ n_{\rm ISM}=1\,{\rm
cm}^{-3}$.  In this work, we investigate the uniform model
jet \citep{Rhoads99}. The shell density and energy is set constant throughout the
shell, and we take it to
correspond to an isotropic spherical shell containing an equivalent
isotropic energy $E_{\rm iso}=10^{51}{\rm ergs}$ and a Lorentz
factor $\gamma=100$. To make the 2D computation feasible, we now
start the simulation with a shell thickness $\delta=10^{14}{\rm cm}$
at a distance $R_{0}=10^{16}{\rm cm}$ from the central engine.  In the
initial setup the shell occupies an annular region,
with half opening angle of the shell equal to $\theta=1^\circ$.
This angle is rather small with respect to those typically
 deduced from modeling of the optical light-curve breaks
 but still in agreement with the most collimated
GRB flows \citep{Bloometal03, Panaitescu05}.
With the choice of a rather narrow jet, we expect the 2D
effects to appear earlier and to be more pronounced with respect
to a spherical shell with the same isotropic equivalent energy.
Note that as a result, this jet like outflow has a decreased
effective energy in the shell $E_{\rm jet}=(\theta^2/2) E_{\rm
iso}$.

The AMR run uses 4 grid levels, taking $400\times 6000$ at level 1,
but with refinement ratios of 2, 4 and 2 between consecutive levels
eventually achieving an effective resolution of $6400 \times 96000$.
The domain is $[0,4]\times[0.3,30]$ in $R_0$ units, as we specifically
intend to model in detail the most dramatic phase of deceleration prior
to the Blanford-McKee evolution.
\begin{figure*}
\begin{center}
\FIG{
{\rotatebox{0}{\resizebox{18cm}{22cm}{\includegraphics{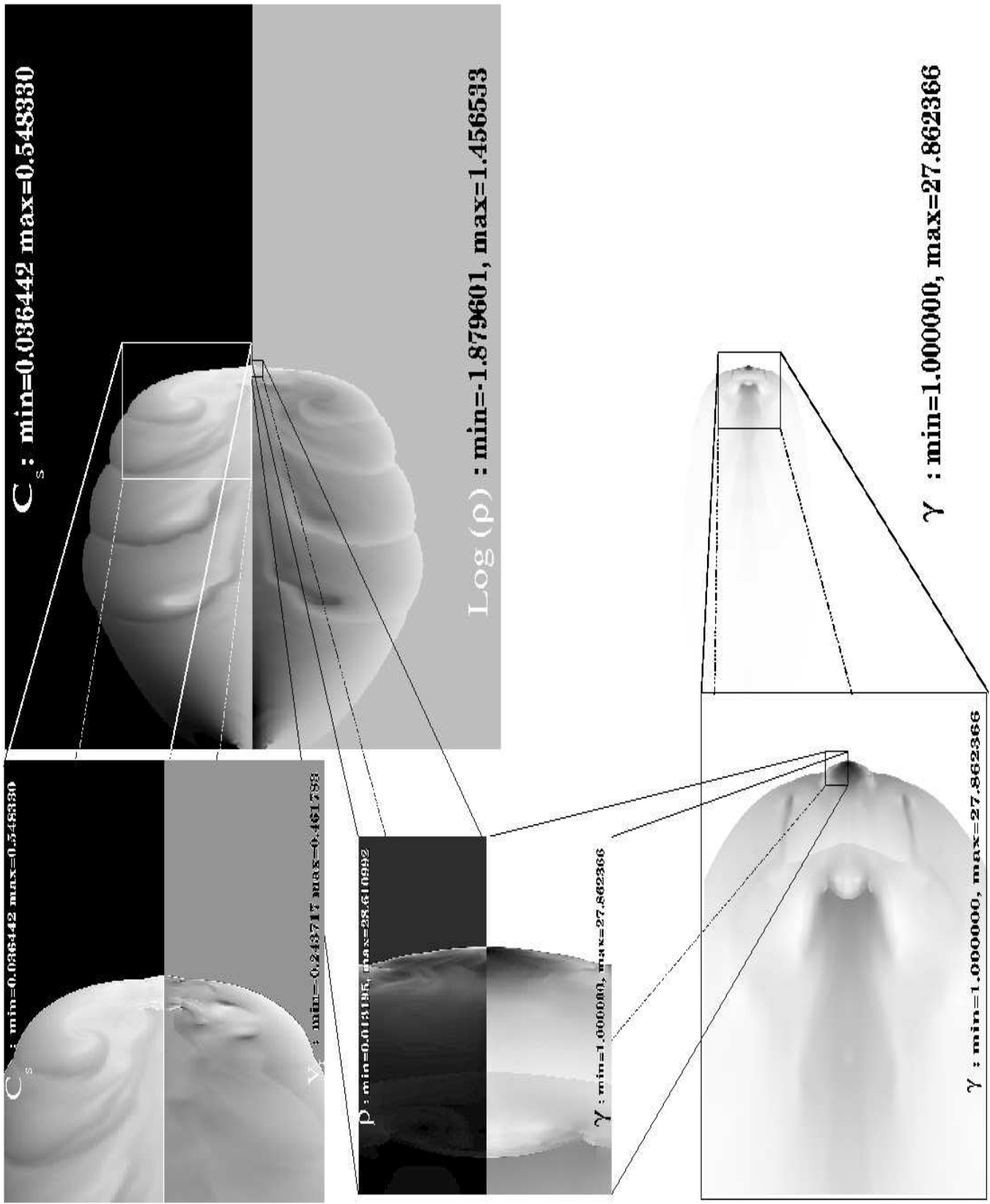}}}}
}
\caption{On the ({\bf top}), the sound speed, logarithm of density
({\bf right}), and Lorentz factor contours for the 2D simulation
({\bf left}). On the ({\bf bottom}), a zoom around on the relativistic
shell, the sound speed and the lateral velocity ({\bf left}), the
Lorentz factor ({\bf right}), and the zoom on the shell, the density and Lorentz factor ({\bf center}) at $t=5\times 10^{6} {\rm s}$.}
\label{LabelFig_11}
\end{center}
\end{figure*}
Note that we initially have $\gamma>1/\theta$, which appears to be
the case for a GRB jet. Beaming effects are invoked to explain the
observed steepening in the decay light curve resultant from the
transition $\gamma>1/\theta$ (indistinguishable from an isotropic
explosion) to $\gamma<1/\theta$ \citep{Panaitescu&Meszaros99,
Panaitescu&Kumar02, Panaitescu05}. With these setup we can
verify from our high resolution simulation whether up to times
corresponding to the transition $\gamma \sim 1/\theta$,  only the
isotropic energy $E_{\rm iso}$ is relevant for the dynamics and the
resulting emission \citep{Piran00, Granot05}.

The initial velocity of the shell is purely
radial. Note that, compared to the 1D isotropic case presented in the
previous section, the 2D
simulation starts with an initial condition containing less energy.
This is done for mere practical reasons: we wish to keep the
computation feasible within two week's execution time on a single
processor. Due to this lower energy content, the deceleration
distance will be smaller  by about one order of magnitude since less
swept-up ISM mass is sufficient
to decelerate the shell. This reduces the need for resolving many
decades of propagation distance as measured in units of the initial
shell thickness.  In fact the 1D equivalent isotropic case with the same
energy shows a sudden decrease of the maximum Lorentz factor of the
decelerating configuration that corresponds to the reverse shock crossing
of the shell at $R\sim 9\times 10^{17} {\rm cm}$. This happens well before
the simulated $3\times 10^{17} {\rm cm}$.

\begin{figure}
\begin{center}
\FIG{
{\rotatebox{0}{\resizebox{8cm}{6cm}{\includegraphics{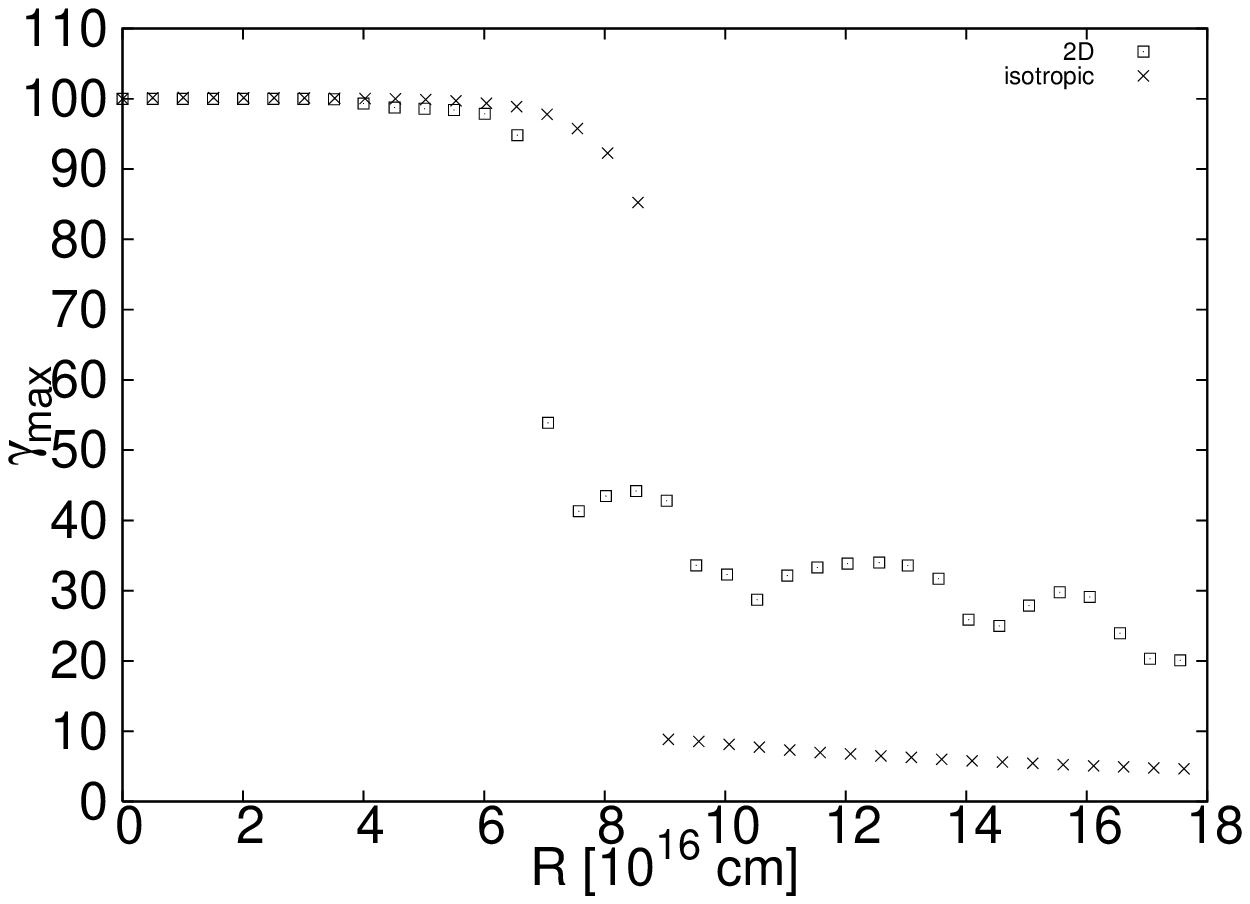}}}} }
\caption{The variation of the maximum Lorentz factor evolution in 1D
and 2D scenarios compared.} \label{LabelFig_12}
\end{center}
\end{figure}

Fig.~\ref{LabelFig_11} shows at the (top), the sound speed contour, and the
density distribution in a logarithmic scale in the  (left), and
the Lorentz factor in the (right). At the bottom, a zoom in the region
 around  the shell is shown in the (left), the sound speed and the lateral velocity,
in the (right) the Lorentz factor, and in the (center) a zoom only on the
shell, the density and Lorentz factor. As in the 1D case, at first the shell
propagates with a constant  maximum Lorentz factor, and this is accompanied
by a weak spread of the  shell. Part of this radial shell
widening in the bottom part of the shell is affected by the creation
of a very low pressure and density region below the shell (also
occurring in the 1D scenario). This near-vacuum state  remains at the
rear part as the shell moves away at the specified Lorentz
factor.  In this 2D simulation, the unshocked shell
also spreads laterally with an initial transverse (horizontal)
velocity, since the shell is launched with a pure radial velocity.
The corresponding maximum initial lateral velocity of the unshocked
shell is $v_{\rm T}=0.0175\,c$. However, the shocked, swept up ISM
matter spreads laterally much faster, due to its high thermal energy
content. Initially, that shocked ISM part spreads with a  comoving velocity of
$v_{\rm T,co}\sim 0.4c$, which is less than the maximum sound speed
allowed by the polytropic equation of state $c/\sqrt{3}$.
Due to this fast sideways expansion of shocked shell and ISM, the mass of the
ISM hit by the shell grows faster than $r^2$. Therefore, the
deceleration of the shell starts earlier than in the isotropic case, see
Fig~\ref{LabelFig_12}. This result implies that the transition
from the phase where $E_{\rm iso}$ is relevant, to the phase where
$E_{\rm jet}$ is relevant in the dynamics takes place when the shocked
ISM and shell start to spread laterally  much faster than what corresponds
to pure radial (ballistic) flow.

In the last part of the shell-ISM deceleration phase, when the
reverse shock has crossed the entire initial shell material, the
lateral velocity of shocked shell material reaches a comoving speed of
$v_{\rm T,co}\sim 0.7c$. This means that we do find that the lateral
velocity can be bigger than the sound speed in the medium which is
in accord with the analytical result of \citet{Sarietal99}. This is
at odds with numerical findings as those found
in~\citet{Cannizzoetal04}, which employ a much reduced resolution as
compared to our AMR results (at low resolution, we do obtain a
reduced lateral spreading velocity). As a result of this fast
lateral spread of the shocked material, distinct differences occur
in the deceleration stages as compared to the isotropic case.
This result is very important, as it shows that the lateral
spreading of the shell is not related only to the Lorentz factor of the
shell but to the type of the reverse shock. In our computation, in an early
phase the reverse shock is Newtonian and the expansion of the shocked shell
part is modest.
However,  in a later phase the reverse shock becomes relativistic
and this leads to faster lateral spreads. However, as the forward shock is
always relativistic,
already in an early stage the shocked ISM spreads with high velocity.
 The overall spreading of the shocked ISM and shocked shell
configuration can, thus, be quite complex and rather more evolved
than the one that semi-analytical models \citep{Rhoads99, Panaitescu&Meszaros99,
Sarietal99, Piran00} predict.

Only that part of the ISM found within the solid angle of
the expanding shell is swept up, and this opening angle changes due
to the spreading effects just discussed. In our 2D simulation, we
found in analogy with the 1D (higher energy) case from above, that
the reverse shock is initially Newtonian, so the thermal energy in
the shocked part of the shell does not increase a lot and its
lateral expansion remains weak for a while. Later on, its  lateral
expansion speed goes up to the $0.7c$ mentioned above, as the
reverse shock becomes relativistic and the material through which
the shocked shell expands laterally has already been brought to
lower densities by the shocked ISM interaction.

The variation of the maximum Lorentz factor is less sudden than in
the equivalent 1D spherical explosion, as quantified in Fig.~\ref{LabelFig_12}.
The part of
the shell most distant from the symmetry axis decelerates before the
more internal part. The shell sweeps up more matter than in the corresponding
isotropic case in the external
parts due to the lateral spreading effects discussed. In fact, in
this simulation we may draw the analogy between the shell
interaction  with the ISM and simulations of relativistic AGN jet
propagation into an external medium.
 As in those cases, the energy is
transferred to the ISM through a bow shock structure.
{ However, our modeled ejected shell representative of a burst in GRBs is
not continually supported by injection of energy at the bottom. As a result, in the (small opening angle) shell there is not really evidence of a clear jet beam as in an AGN jet.
In this case the interaction shell-ISM is dominated primarily by forward, reverse shock pair, and contact
discontinuity in between.}
 The
changing 2D structure of this shock leads to differences in the
shocked ISM mass loaded on to the shell. As stated earlier, only a
fraction of the ISM within the  opening angle of the shell is swept
up, and an important part of ISM matter gets pushed away laterally
as the thin radially confined shell advances.
 The resulting behavior is clearly
influenced by these 2D effects, and is the reason why the maximum
Lorentz factor of the configuration starts to decrease earlier than
in the 1D scenario. In an isotropic scenario all swept up mass of
the ISM remains in front of the initial shell where most of the
energy of the shell is continually transferred to shocked ISM.
However, in the jet-like explosion, the shocked ISM and shell expand
laterally leading to interaction with more ISM material, but the
main part of this ISM material gets deflected about the shell. As
seen in Fig~\ref{LabelFig_11}, we find rather complex flow patterns
trailing the thin shell. Hence, the mass accreted on to the shell in
fact decreases as compared to the equivalent local isotropic
scenario. Therefore, less energy is transferred continually to the
shocked ISM.

 One interesting characteristic
of the maximum Lorentz factor of the beamed shell at radii
$R>7\times 10^{16} {\rm cm}$
 is the ``bumps'' that it shows as function of radius. These modulations of
$\gamma_{\rm max}$ are a result of rapid internal motions of the decelerating
configuration caused by the complex shell-ISM interaction.
In view to the very rich and unexpected early afterglow phenomenology revealed
by the {\it SWIFT} satellite (see, for example, \citet{Zhangetal06}), it is
interesting to study whether these proper motions can cause a
significant modulation in the emitted radiation expected from these flows.

In this simulation, we find no real indication of a strong change in
lateral spreading of the shell when the Lorentz factor drops down to
$\gamma_{c}=1/\theta\sim 57$ \citep{Rhoads99, Panaitescu&Meszaros99}.
In fact, an important change is produced later, when the Lorentz
factor of the shell becomes smaller than $30$, while the lateral
velocity of the shell reaches $v_{\rm T,co}\sim 0.7c$. As pointed out, this
coincides with the time when the reverse shock became relativistic
and almost crossed the initial shell entirely. After this phase of
rapid lateral spread, we find that the spread out shell decelerates
faster, as it accumulates more matter (Fig.~\ref{LabelFig_11}).
 More simulations with different values for the shell opening angle and
thickness are to investigate how these parameters affect the phase  of the
deceleration where the lateral spreading of the shell becomes important.

\section{Conclusions}
In this paper, we presented and applied the AMRVAC code in its extension to relativistic hydrodynamics.
The adaptive mesh refinement is particularly useful for simulating highly relativistic flow dynamics.
We always used the robust TVDLF sheme, and this shock-capturing method together with high effective resolution
delivers numerical results that can rival or even improve other
high order methods. As is well-known, difficulties
in special relativistic hydrodynamic simulations result from the non linear coupling
between different components of the velocity by the Lorentz factor and also the coupling between
inertial and thermodynamics. We demonstrated that
Adaptive Mesh Refinement (AMR) is then very useful to resolve the associated very thin structures properly.
We tested the code ability with stringent recent
test problems collected from the astrophysical literature, including 1D and 2D
shock tube problems, an ultrarelativistic flow reflecting of a wall, a relativistic
variant of a forward-faced
reflecting step, and 2D astrophysical jet propagation.

We used the fireball model to investigate 1D and 2D afterglow phases
in GRBs. In 1D, we examined the evolution of a cold relativistic
shell with a Lorentz factor of $100$ in uniform medium. In this
simulation we discussed details of the internal structure of the
evolving shell-ISM ejecta  and compare them with analytical estimates.
We followed the evolution of this isotropic explosion  almost
all the way into the classical Sedov phase. At all times, we resolve the various regions
that characterize this interaction. We quantified and discussed the
precise deceleration of the relativistic shell. When most of the
initial energy of the shell is transferred to swept-up shocked ISM
(this occurs at the
forward shock), the deceleration of shocked ISM is eventually well
described by the relativistic Blandford-McKee self-similar solution.
Hence the Lorentz factor of the forward shock decreases as
$R^{-3/2}$.

We investigate also the afterglow phase for a  beamed 2D shell. In this
model, we discussed analogies and important differences with the 1D
model. The interaction of a confined relativistic shell with the ISM
is characterised by the appearance of a bow shock. We showed how ISM
material is laterally pushed out, thus decreasing the amount of
accumulated matter in front of the shell near the axis. The part of the shell furthest
away from the axis decelerates then faster than in a 1D spherical case.
Although the deceleration of the shell starts early as compared to an equivalent isotropic
case,
the deceleration of the inner part of the shell is slow due to the
weak accreted ISM matter in front of the shell. The thermal
energy of the shocked ISM increases and induces a lateral spread of
this shocked ISM. We have shown with a high resolution simulation of
jet-like GRB models in their afterglow phase that this lateral
expansion goes through various phases.

First, the shell spreads only
with its initial lateral velocity until it accretes enough ISM
matter. In this phase, the shocked ISM spreads laterally
with a velocity near the sound speed. However, the reverse
shock  propagates in a Newtonian fashion through the shell, thus having a small efficiency
in the conversion of the kinetic energy of the shell to thermal energy, hence the
expansion of the shocked shell is still weak. Only in a later phase
when the reverse shock becomes relativistic, the lateral expansion
of the shocked shell increases drastically and reaches a high
velocity $v_{T}\sim 0.7 c$.
The transition from slow to fast lateral spreading of the shell is
thus related to the transition from Newtonian to relativistic reverse
shock propagation. However, as the forward shock is always relativistic the
shocked ISM spreads laterally faster.

 The 2D simulation has revealed rapid internal motions in the
 decelerating configuration. It is possible that these motions result
in modulations in the afterglow emission.
In future work, we intend to use these and similar simulation results to compute
their predictions for the precise afterglow spectral evolution.

\section*{acknowledgements}
We acknowledge financial support from the Netherlands Organization for
Scientific Research, NWO-E grant 614.000.421,
and computing resources supported by
NCF. Part of the computations made use of the VIC cluster at K.U.Leuven. Part of this
research was supported by European FP5 RTN "Gamma Ray Burst: An Enigma and a
Tool".


\begin{thebibliography}{99}
\bibitem[\protect\citeauthoryear{Aloy et al.}{2000}]{Aloy00}Aloy, M. A.,
 M{\"u}ller, E., Ib\'anez, J. M., Mart\'i, J. M., MacFadyen, A., 2000, ApJ,
531, L119
\bibitem[\protect\citeauthoryear{Arons}{2004}]{Arons04}Arons, J., 2004, Advances in Space Research,
33, 466
\bibitem[\protect\citeauthoryear{Barthelmy et al.}{2005}]{Barthelmy05}Barthelmy S. D., et al.,2005, ApJ, 635, L133
\bibitem[\protect\citeauthoryear{Begelman \& Cioffi}{1989}]{Begelman&Cioffi89}
Begelman, M. C., Cioffi, D. F., 1989, ApJ, 345, L21
\bibitem[\protect\citeauthoryear{Berger et al.}{2000}]{Berger00}Berger E., et al. 2000, ApJ, 545, 56
\bibitem[\protect\citeauthoryear{Berger et al.}{2003}]{Bergeretal03}Berger, E., Kulkarni, S. R.,
 Frail, D. A., 2003, ApJ, 590, 379

\bibitem[\protect\citeauthoryear{Bergmans et al.}{2004}]{Bergmansetal04} J. Bergmans, R. Keppens, D.E.A.
van Odyck and A. Achterberg, Lecture Notes in Computer Science and
Engineering,
 Proceedings of Chicago Workshop on Adaptive Mesh Refinement Methods Sept. 3-5 2003
\bibitem[\protect\citeauthoryear{Blandford \& McKee}{1976}]{Blandford&McKee76}
Blandford, R. D., \& McKee, D. G., 1976, Physics of Fluids, 19,1130
\bibitem[\protect\citeauthoryear{Bloom et al.}{2003}]{Bloometal03}Bloom, J. S., Frail, D. A., Kulkarni, S. R.,
 2003, ApJ, 594, 674
\bibitem[\protect\citeauthoryear{Cheng et al.}{2001}]{Chengetal01}Cheng, K. S., Huang, Y. F.,
Lu. T., 2001, MNRAS, 325, 599

\bibitem[\protect\citeauthoryear{Cannizzo et al.}{2004}]{Cannizzoetal04}Cannizzo, J. K., Gehrels N.,
Vishniac E. T., 2004, ApJ, 601, 380


\bibitem[\protect\citeauthoryear{Cenko et al.}{2006}]{Cenkoelal06}Cenko, S. B. et al., 2006,
astro-ph/0608183

\bibitem[\protect\citeauthoryear{Chevalier \& Li}{2000}]{Chevalier&Li00} Chevalier, R. A., Li
Z.-Y, 2000, ApJ, 536, 195
\bibitem[\protect\citeauthoryear{Chevalier et al.}{2004}]{Chevalieretal04} Chevalier, R. A., Li Z.-Y,
Fransson, C., 2004, ApJ, 606, 369

\bibitem[\protect\citeauthoryear{Chiang \& Dermer}{1999}]{Chiang&Dermer99}Chiang J., Dermer C. D.,
1999, ApJ,  512, 699
\bibitem[\protect\citeauthoryear{Corbel}{2004}]{Corbel04}Corbel, X-ray Timing 2003: Rossi and Beyond. AIP Conference Proceedings, Vol. 714, held 3-5 November, 2003 in Cambridge, MA. Edited by Philip Kaaret, Frederick K. Lamb, and Jean H. Swank. Melville, NY: American Institute of Physics,
2004, 127

\bibitem[\protect\citeauthoryear{Costa et al.}{1997}]{Costa97}Costa, E., et al., 1997, Nature, 387, 783

\bibitem[\protect\citeauthoryear{Covino et al.}{1999}]{Covinoetal99}Covino, S.et al. , 1999, A\&A, 348, L1
\bibitem[\protect\citeauthoryear{Dar \& De R\'ujula}{2004}]{Dar&Rujula04}Dar, A., De Rujula, A. 2004, Physics
Reports, 405, 203
\bibitem[\protect\citeauthoryear{Del Zanna \& Bucciantini}{2002}]{DelZanna&Bucciantini02}Del Zanna, L., Bucciantini, N., 2002,
A\&A, 390, 1177
\bibitem[\protect\citeauthoryear{Donaghy}{2006}]{Donaghy06}Donaghy, T. Q., 2006, ApJ, 645, 436
\bibitem[\protect\citeauthoryear{Emery}{1968}]{Emery1968}Emery, A. E., 1968, J. Comput. Phys., 2, 306
\bibitem[\protect\citeauthoryear{Ferrari}{1998}]{Ferrari98}Ferrari, A., 1998,
Annu. Rev. Astron. Astrophys., 36, 539

\bibitem[\protect\citeauthoryear{Frail et al.}{1997}]{Frail97}Frail, D. A., Kulkarni, S. R.,
Nicastro, S. R., Feroci, M., Taylor, G. B., 1997, Nature, 389, 261
\bibitem[\protect\citeauthoryear{Frail et al}{2001}]{Frailetal01}Frail, D. A., et al.,  2001, ApJ, 562, L55


\bibitem[\protect\citeauthoryear{Frail et al.}{2006}]{Frailetal06} Frail, D. A. et al., 2006, ApJ, 646, L99
\bibitem[\protect\citeauthoryear{Galama et al.}{1997}]{Galama97}Galama, T.,
et al. , 1997, Letters to Nature,  387,  479
\bibitem[\protect\citeauthoryear{Galama et al.}{1998}]{Galamaetal98}
Galama, T. J., et al., 1998, ApJL, 500, L97
\bibitem[\protect\citeauthoryear{Gorosabel et al.}{}{2006}]{Gorosabeletal06}Gorosabel, J. ,et al. , 2006 astr-ph/0603100

\bibitem[\protect\citeauthoryear{Granot et al}{2001}]{Granot01}Granot, J., Miller, M., Piran, T., Suen, W. M., Hughes, P. A., 2001, Gamma-Ray Bursts in the Afterglow Era: Proceedings of the International Workshop Held in Rome, Italy, 17-20 October 2000, ESO ASTROPHYSICS SYMPOSIA. ISBN 3-540-42771-6. Edited by E. Costa, F. Frontera, and J. Hjorth. Springer-Verlag, 2001,  312
\bibitem[\protect\citeauthoryear{Granot}{2005}]{Granot05}Granot, J, 2005, Triggering relativistic jets,
RevMexAA,astro-ph/0610379
\bibitem[\protect\citeauthoryear{Greiner et al.}{2003}]{Greineretal03}Greiner, J. et al. 2003, Nature,
426, 157
\bibitem[\protect\citeauthoryear{Katz}{1994}]{Katz94}Katz, J. I., 1994, ApJ, 432, 107L
\bibitem[\protect\citeauthoryear{Keppens et al.}{2003}]{Keppensetal03}Keppens, R., Nool, M., T\'oth, G.,
Goedbloed, J. P., 2003,  153, 317
\bibitem[\protect\citeauthoryear{Kobayashi et al.}{1999}]{Kobayashietal99}Kobayashi S., Piran, T.,
Sari, R., 1999, ApJ, 513, 669
\bibitem[\protect\citeauthoryear{Kobayashi \& Sari}{2000}]{Kobayashi&Sari00}Kobayashi S., Sari, R., 2000,
ApJ, 542, 819

\bibitem[\protect\citeauthoryear{Kulkarni et al.}{1999}]{Kulkarni99}Kulkarni, S. R. et al., 1999, ApJ,
 522, 97L

\bibitem[\protect\citeauthoryear{Kumar \& Panaitescu}{2000}]{Kumar&Panaitescu00}Kumar, P.,Panaitescu, A. 2000, ApJ, 541, L9

\bibitem[\protect\citeauthoryear{Kumar \& Granot}{2003}]{Kumar&Granot03}Kumar, P., Granot ,
2003,  ApJ, 591, 1075
\bibitem[\protect\citeauthoryear{Lazzati et al.}{2004}]{Lazzatietal04}Lazzati, D., et al. , 2004,
A\&A, 422,  121L
\bibitem[\protect\citeauthoryear{Lucas-Serrano et al.}{2004}]{Lucas-Serranoetal04}Lucas-Serrano, A.,
Font, J, A., Iban\'a\~{n}ez J. M, and Marti, J. M., 2004, A\&A, 428,
703
\bibitem[\protect\citeauthoryear{Marti \& M\~{u}ller}{1994}]{Marti&Muller1994}Marti, J. M.,
M\"{u}ller E., 1994,  J. Fluid Mech., 258, 317
\bibitem[\protect\citeauthoryear{Marti et al.}{1997}]{Martietal97}Marti, J. M., M\"{u}ller E.,
Font J. A., Iban\'a\~{n}ez J. M, and Marquina A., 1997, APJ, 479,
151
\bibitem[\protect\citeauthoryear{Marti \& M\~{u}ller}{2003}]{Marti&Muller2003}Marti, J. M.,
M\"{u}ller E., 2003,  Living Rev. Relativity, 6, 7

\bibitem[\protect\citeauthoryear{M\'esz\'aros \& Rees}{1992}]{Meszaros&Rees92}M\'esz\'aros P., Rees M. J.,
 1992, MNRAS, 258, 41P

\bibitem[\protect\citeauthoryear{M\'esz\'aros \& Rees}{1997}]{Meszaros&Rees97}M\'esz\'aros P., Rees M. J.,
 1997, ApJ, 476, 232

\bibitem[\protect\citeauthoryear{M\'esz\'aros}{2006}]{Meszaros06}M\'esz\'aros, P., Meszaros, P.,
2006, Rep. Prog. Phys., 69, 2259

\bibitem[\protect\citeauthoryear{Metzger}{1997}]{Metzger97}Metzger, M. R., Djorgovski, S. G., Kulkarni,
S. R., Steidel, C. C., Adelberger, K. L., Frail, D. A., Costa, E.,
Frontera, F., 1997, Nature, 387, 879

\bibitem[\protect\citeauthoryear{Mimica et al.}{2006}]{Mimicaetal06}
Mimica, P. Aloy, M. A., M\"uller E.,
astro-ph/0611765
\bibitem[\protect\citeauthoryear{Mignone et al.}{2005}]{Mignoneetal05}Mignone, A., Plewa, T., Bodo, G., 2005, ApJS, 160, 199
\bibitem[\protect\citeauthoryear{Oren et al.}{2004}]{Orenetal04}Oren, Y., Nakar, E., Piran, T., 2004, MNRAS, 353, L35
\bibitem[\protect\citeauthoryear{Panaitescu et al.}{1997}]{Panaitescuetal97}Panaitescu, A., Wen, A.,
Laguna, P., M\'esz\'aros, P., 1997, ApJ, 482, 942
\bibitem[\protect\citeauthoryear{Panaitescu}{2005}]{Panaitescu05}Panaitescu A., 2005, MNRAS,
363, 1409
\bibitem[\protect\citeauthoryear{Panaitescu \& M\'esz\'aros}{1999}]{Panaitescu&Meszaros99}
Panaitescu, A., M\'esz\'aros, P., 1999, ApJ, 526, 707
\bibitem[\protect\citeauthoryear{Panaitescu \& Kumar}{2002}]{Panaitescu&Kumar02}Panaitescu, A., Kumar
2002, ApJ, 571, 779
\bibitem[\protect\citeauthoryear{Panaitescu \& Kumar}{2003}]{Panaitescu&Kumar03}Panaitescu, A., Kumar
2003, ApJ, 592, 390
\bibitem[\protect\citeauthoryear{Piner et al.}{2003}]{Pineretal03}
Piner, B. G., Unwin, S. C., Wehrle, A. E., Zook, A. C., Urry, C. M.,
\& Gilmore, D. M. 2003, ApJ, 588, 716
\bibitem[\protect\citeauthoryear{Piran}{2000}]{Piran00} Piran, T., 2000,  Physics Reports, 333, 529
\bibitem[\protect\citeauthoryear{Piran}{2005}]{Piran05}Piran, T., 2005,  Rev. Mod. Phys., 76, 1143

\bibitem[\protect\citeauthoryear{Piro et al.}{1998}]{Piroetal98}Piro et al., 1998, A\&A, 331, L41
\bibitem[\protect\citeauthoryear{Pons et al.}{2000}]{Ponsetal00}Pons J. A., Marti, J. M.,
 M\"{u}ller E., 2000, J. Fluid Mech., 422, 125
\bibitem[\protect\citeauthoryear{Rhoads}{1993}]{Rhoads93}Rhoads, J. E., 1993, ApJ, 591, 1097
\bibitem[\protect\citeauthoryear{Rhoads}{1997}]{Rhoads97}Rhoads, J. E., 1997, ApJ, 478, L1
\bibitem[\protect\citeauthoryear{Rhoads}{1999}]{Rhoads99}Rhoads, J. E., 1999, ApJ, 525, 737
\bibitem[\protect\citeauthoryear{Sahu et al.}{1997}]{Sahu97}Sahu, K. C., Livio, M., Petro, L., Macchetto, F. D.,
 van Paradijs, J., Kouveliotou, C., Fishman, G. J., Meegan, C. A., Groot, P. J.,
 Galama, T., 1997, Nature, 387, 476
\bibitem[\protect\citeauthoryear{Sari \& Piran}{1995}]{Sari&Piran95}Sari R.,  Piran T., 1995, ApJL, 455, L143
\bibitem[\protect\citeauthoryear{Sari et al.}{1998}]{Sarietal98} Sari, R., Piran, T, \& Narayan, R., 1998, ApJL, 497, L17
\bibitem[\protect\citeauthoryear{Sari \& Piran}{1999}]{Sari&Piran99}Sari R., Piran T., 1999, ApJL, 517, L109
\bibitem[\protect\citeauthoryear{Sari et al.}{1999}]{Sarietal99}Sari R., Piran T., \& Halpem, J. P.,
1999, ApJ, 519, L17

\bibitem[\protect\citeauthoryear{Shemi \& Piran}{1990}]{Shemi&Piran90}Shemi, A., Piran, T.,
 1990, ApJ, 365, L55

\bibitem[\protect\citeauthoryear{Soderberg \& Ramirez-Ruiz}{2001}]{Soderberg&Ramirez-Ruiz01}
Soderberg A. M., Ramirez-Ruiz E., 2001, AIP conf. Proc. 662:
Gamma-Ray Burst and Afterglow Astronomy 2001, pp172-175

\bibitem[\protect\citeauthoryear{Stanek et al.}{1999}]{Staneketal99} Stanek,  K. Z., Garnavich, P. M.,
 Kaluzny, J., Pych, W., Thompson, I. , 1999, ApJ, 522, 39L
\bibitem[\protect\citeauthoryear{T\'oth \& Odstr\v
    cil}{1996}]{Toth&Odstrvcil} G. T\'oth and D. Odstr\v cil,  1996,
   J. Comput. Phys., 128, 82.
\bibitem[\protect\citeauthoryear{Van der Holst \& Keppens}{2006}]{holstkep06}van der Holst B., Keppens R., 2006, J. Comput. Phys., submitted
\bibitem[\protect\citeauthoryear{Van Marle et al.}{2006}]{VanMarleetal06}Van Marle A. J.,
N. Langer, A. Achterberg, G. Garcia-Segura, 2006, astro-ph/0605698
\bibitem[\protect\citeauthoryear{Van Paradijs et al.}{1997}]{vanParadijsetal97}Van Paradijs, J., et al. ,
1997, Nature, 386, 686

\bibitem[\protect\citeauthoryear{Vietri}{1997}]{Vietri97}Vietri, M. , 1997, ApJ, 478, L9

\bibitem[\protect\citeauthoryear{Wijers et al.}{1999}]{Wijersetal99}Wijers, R. A. et al., 1999, ApJ,
 523, L33
\bibitem[\protect\citeauthoryear{Wijers}{1997}]{Wijers97}Wijers, R., 1997, Nature, 393, 13
\bibitem[\protect\citeauthoryear{Woodward \& Colella}{1984}]{Woodward&Colella1984}Woodward, P.,
 Colella, P., 1984, J. Comput. Phys., 54, 115
\bibitem[\protect\citeauthoryear{Woods \& Loeb}{1995}]{Woods&Loeb95}Woods E., Loeb A., 1995,
ApJ, 453, 583
\bibitem[\protect\citeauthoryear{Zhang et al.}{2006}]{Zhangetal06} Zhang, B., et al., 2006, ApJ, 642, 354
\bibitem[\protect\citeauthoryear{Zhang \& MacFadyen}{2006}]{Zhang&MacFadyen06}Zhang W.,
MacFadyen A. I. , 2006, ApJS, 164, 255
\end{thebibliography}
\end{document}